\documentclass[%
reprint,
showkeys,
superscriptaddress,
preprintnumbers,
nofootinbib,
amsmath,amssymb,
aps,
prd,
floatfix,
]{revtex4-2}
\usepackage{afterpage}
\usepackage{subfigure}
\usepackage{rotating} 
\usepackage{booktabs}
\usepackage{graphicx}
\usepackage{dcolumn}
\usepackage{bm}
\usepackage{float}
\usepackage{amsmath}
\usepackage{natbib}
\usepackage{hyperref}

\usepackage{xcolor}
\newcommand{\reply}[1]{{ #1}}

\begin{document}



\preprint{H.~Song \& B.-Q.~Ma, \href{https://doi.org/10.1103/PhysRevD.111.103015}{Phys. Rev. D 111 (2025) 103015}}

\title{Monte Carlo simulation of GRB data to test Lorentz-invariance violation}

\author{Hanlin Song}
\affiliation{School of Physics, Peking University, Beijing 100871, China}

\author{Bo-Qiang Ma}\email{mabq@pku.edu.cn}
\thanks{corresponding author.}
\affiliation{School of Physics, Zhengzhou University, Zhengzhou 450001, China}
\affiliation{School of Physics, Peking University, Beijing 100871, China}	
\affiliation{Center for High Energy Physics, Peking University, Beijing 100871, China}

\begin{abstract}
Lorentz-invariance violation (LV) at energy scales approaching the Planck regime serves as a critical probe for understanding quantum gravity phenomenology. Astrophysical observations of gamma-ray bursts (GRBs) present a promising avenue for testing LV-induced spectral lag phenomena; however, interpretations are complicated by degeneracies between LV effects and intrinsic emission delays. This study systematically investigates three competing time delay models: Model A (LV delay combined with a constant intrinsic delay), Model B (energy-dependent intrinsic delay without LV), and Model C (LV delay combined with energy-dependent intrinsic delay). We utilize mock GRB datasets generated under distinct delay mechanisms and employ Bayesian parameter estimation on simulated observations of 10 GRBs. Our findings demonstrate that Model C consistently recovers input parameters across all datasets, including those designed to mimic LV-dominated (Model A) and intrinsic delay-dominated (Model B) scenarios. In contrast, Models A and B struggle to reconcile data generated under alternative mechanisms, particularly when confronted with high-energy TeV photons from GRB 190114C and GRB 221009A. Our analysis confirms that the incorporation of energy-dependent intrinsic delays in Model C is essential for establishing robust LV constraints, effectively resolving prior ambiguities in the interpretation of multi-GeV and TeV photon emissions. The results validate Model C as a generalized framework for future LV searches, yielding a subluminal LV scale of \(E_{\rm LV} \simeq 3 \times 10^{17}\) GeV based on realistic datasets, including 14 Fermi-LAT multi-GeV photons from eight GRBs, a 99.3 GeV photon from GRB 221009A observed by Fermi-LAT, a TeV photon from GRB 190114C detected by MAGIC, and a multi-TeV photon from GRB 221009A recorded by LHAASO.  These findings are consistent with earlier constraints derived from Fermi-LAT datasets. This work underscores the necessity for joint modeling of LV and astrophysical emission processes in next-generation LV studies utilizing observatories such as LHAASO and CTA.
\end{abstract}

\maketitle

\section{Introduction}
Lorentz invariance, a foundational symmetry of modern physics underpinning both special relativity and quantum field theory, has undergone rigorous scrutiny in recent decades due to mounting theoretical interest in potential Lorentz-invariance violation (LV) at extreme energy scales. 
Numerous quantum gravity frameworks—including string theory \cite{Amelino-Camelia:1996bln, Amelino-Camelia:1997ieq, Ellis:1999rz, Ellis:1999uh, Ellis:2008gg, Li:2009tt, Li:2021gah, Li:2021eza}, loop quantum gravity \cite{Gambini:1998it, Alfaro:1999wd, Li:2022szn}, and doubly-special relativity \cite{Amelino-Camelia:2002cqb, Amelino-Camelia:2002uql, Amelino-Camelia:2000stu}—predict deviations from Lorentz symmetry near the Planck energy scale, $E_{\rm P} \equiv \sqrt{\hbar c^5/G} \simeq 1.22 \times 10^{19}$ GeV. 
These theoretical developments have spurred extensive efforts to detect LV through astrophysical observations \cite{He:2022gyk, AlvesBatista:2023wqm}.

A phenomenological approach to parametrizing LV effects involves modifying the photon dispersion relation. Expanding the energy-momentum relation as a Taylor series, the leading-order correction for a photon with energy $E$ and momentum $p$ takes the form
\cite{He:2022gyk,Xiao:2009xe} 
\begin{equation}
E^2=p^2c^2\left[1-s_n\left(\frac{pc}{E_{\mathrm{LV},n}}\right)^n\right],
\end{equation}
where $s_n \equiv \pm 1$ determines whether high-energy photons propagate superluminally ($s_n = - 1$) or subluminally ($s_n = + 1$) relative to their low-energy counterparts and 
$E_{{\rm LV},n}$ represents the characteristic LV energy scale for the 
$n$th-order correction. The corresponding velocity modification derived from $v = \partial E / \partial p$ becomes
\begin{equation}
v(E) = c\left[1 - s_n\frac{n+1}{2}\left(\frac{pc}{E_{\mathrm{LV},n}}\right)^n\right].
\end{equation}

Direct detection of such minuscule velocity differences in terrestrial experiments remains impractical due to the Planck-scale suppression of 
$E_{{\rm LV},n}$. However, Amelino-Camelia \textit{et al}. \cite{Amelino-Camelia:1996bln, Amelino-Camelia:1997ieq} proposed that cosmological gamma-ray bursts (GRBs) could amplify these effects: the vast propagation distances of GRB photons allow velocity discrepancies to accumulate into measurable arrival time differences between high- and low-energy photons. 
With the cosmic expansion taken into account, the LV-induced time delay for a source at redshift $z$ is expressed as \cite{Jacob:2008bw, Zhu:2022blp}
\begin{equation}
\label{lvdelay}
\Delta t_{\mathrm{LV}}=s_{n}\frac{1+n}{2H_{0}}\frac{E_{\mathrm{h,o}}^{n}-E_{\rm l,o}^{n}}{E_{\mathrm{LV},n}^{n}}\int_{0}^{z}\frac{(1+z')^{n}\mathrm{d}z'}{\sqrt{\Omega_{\mathrm{m}}(1+z')^{3}+\Omega_{\Lambda}}},
\end{equation}
where $E_{\rm h,o}$ and $E_{\rm l,o}$ are observed energies of high- and low-energy photons, $z$ is the redshift of GRB, $H_0$ is the Hubble constant, and $\Omega_{\mathrm{m}}$ and $\Omega_{\Lambda}$ are matter density parameter and dark energy density parameter of the $\Lambda {\rm CDM}$ model. The low-energy $E_{\rm l,o}$ is omitted during the calculation for the tiny value.

The observed time delay $\Delta t_{\mathrm{obs}}$ comprises both LV and intrinsic emission components,
\begin{equation}
\label{obsdelay}
\Delta t_{\mathrm{obs}} = \Delta t_{\mathrm{LV}} + (1+z)\Delta t_{\mathrm{in}},
\end{equation}
where $\Delta t_{\mathrm{in}}$ corresponds to source-frame emission delays between energy bands. 
Previous works~\cite{Shao:2009bv, Zhang:2014wpb, Xu:2016zxi, Xu:2016zsa, Liu:2018qrg,Zhu:2021pml, Zhu:2021wtw, Zhu:2022usw} treated the intrinsic time delay term as a common constant for all GRB photons. For example, Refs.~\cite{Xu:2016zxi,Xu:2016zsa} analyzed 14 multi-GeV photons from eight GRBs by LAT of Fermi Gamma-ray Space Telescope (Fermi-LAT), and the results suggest that $n=  1$, $s_n = + 1$, $E_{\rm LV,1} \simeq 3.60 \times 10^{17}$ GeV, and $\Delta t_{\rm in} = -10.7$ s for the mainline photons (see Fig.~2 of Ref. \cite{Xu:2016zsa}). These findings suggest the subluminal aspect of cosmic photon Lorentz violation. 

Recent work in Ref. \cite{Song:2024and} critically reexamined these assumptions by introducing three distinct models: 
\textbf{Model A} (replicating prior approaches with LV delay + constant  $\Delta t_{\mathrm{in}}$); 
\textbf{Model B} (excluding LV, i.e., choosing $\Delta t_{\mathrm{LV}} =0$, but introducing energy-dependent $\Delta t_{\mathrm{in}}$); 
\textbf{Model C} (combining LV delay with energy-dependent intrinsic delays). 

Notably, Model C unifies Model A and Model B and achieves consistency with both historical Fermi-LAT data and new TeV-photon detections from GRB 221009A (99.3 GeV by Fermi-LAT \cite{Lesage:2023vvj} and 12.2 TeV by LHAASO \cite{LHAASO:2023lkv}) and GRB 190114C (1.07 TeV by MAGIC \cite{MAGIC:2019lau}), yielding 
$E_{\rm LV}\sim 3\times 10^{17}$~GeV \cite{song_and_ma_ApJ, Song:2025qej}. In contrast, Models A and B fail to reconcile TeV observations with multi-GeV data \cite{Song:2025qej}, highlighting the necessity of energy-dependent intrinsic delays in LV analyses.

This work extends these investigations through systematic simulations of three mock GRB datasets (10 bursts each), generated under the delay mechanisms of Models A, B, and C. Employing Bayesian parameter estimation \cite{Song:2024and}, we demonstrate the following:\\
\noindent
(i) Parameter recovery remains robust across all models when analyzed self-consistently;\\
\noindent
(ii) Model C exhibits superior generality in accommodating diverse delay mechanisms;\\
\noindent
(iii) Energy-dependent intrinsic delays are essential for reliable LV constraints.

Our results solidify Model C as the preferred framework for future LV searches with next-generation observatories.

\section{Mock data generation}

Physics is an experimental science, and it is important for theories to be based on data from observations rather than from preassumptions. In our study, we analyze the arrival time difference between high- and low-energy photons from GRBs to investigate Lorentz-invariance violation. The observed time difference may be attributed to two factors: the time delay caused by light-speed variation (or Lorentz violation) for photons traveling from GRBs to our observer, as derived from standard cosmology~\cite{Jacob:2008bw} or Finsler geometry~\cite{Zhu:2022blp}, respectively; and the intrinsic time delay in the GRB source frame, which may vary depending on the energy of the photon.

Our approach to determining the intrinsic photon emission time is established within a general framework devoid of any arbitrary assumptions, with the parameters derived exclusively through data fitting. It is a fundamental mathematical principle that any analytical function can be represented by a Taylor expansion within a certain range. Consequently, we can express the general relationship between the intrinsic photon emission time 
$\Delta t_{\rm{in}}$ and energy $E$ as a Taylor expansion involving terms up to $E^n$,
$$
\Delta t_{\rm{in}}=\Delta t_{\rm{in},c}+\alpha E+\beta E^2+\gamma E^3+\cdots,
$$
where $\Delta t_{\rm{in},c}=\mu$ represents the constant term. By incorporating additional $E^2$ and $E^3$ terms in the intrinsic photon emission time, 
$\Delta t_{\rm{in}}$ can be regarded as a Taylor expansion of an analytical function within a certain level of precision. Consequently, our model integrating both the Lorentz violation term and the energy dependence of intrinsic emission time can be viewed as a specific instance of this general framework, eliminating the necessity for arbitrary assumptions regarding the parameters (i.e., the coefficients) and deriving them solely through data fitting. Moreover, it has been demonstrated that by applying the aforementioned general expression of $\Delta t_{\rm{in}}$ up to the $E^3$ term to fit the data, it is notable that for both scenarios involving 14 Fermi data and 14 Fermi data combined with three new Fermi+MAGIC+LHAASO data, the coefficients $\beta$ and $\gamma$ converge toward 0 through data fitting \cite{song_and_ma_ApJ}.    

As a first approximation, we can take the intrinsic emission time with energy-dependence as~\cite{Song:2024and}
\begin{equation}
    \Delta t_{\rm{in}} = \Delta t_{\rm in, c} + \alpha E_{\rm h,s},
\end{equation}
where $\Delta t_{\rm in, c}$ is a common constant term and $E_{\rm h,s}$ is the source frame energy of high-energy photon with $\alpha$ being the coefficient. 
This model of intrinsic emission time can be considered as a rational framework capable of accommodating all GRB photon data for the analysis of Lorentz violation by incorporating the energy dependence of intrinsic emission time.

Then, three distinct models in relating the observed time delay $\Delta t_{\mathrm{obs}}$ with the Lorentz violation time delay $\Delta t_{\mathrm{LV}}$ and the intrinsic time delay $\Delta t_{\rm in}$ are cataloged as follows~\cite{Song:2024and}:\\
\begin{itemize}
\item {\bf{Model A}}: with the inclusion of the Lorentz violation term $\Delta t_{\mathrm{LV}}$ by taking $\Delta t_{\rm in}$ as a common constant term for all high-energy photons under study
\begin{equation}
\label{obsdelayA}
\Delta t_{\mathrm{obs}}=\Delta t_{\mathrm{LV}}+(1+z)\Delta t_{\mathrm{in},c};
\end{equation}
\item {\bf{Model B}}: taking $\Delta t_{\rm in}$ as the general form with a common constant term and a linear-type energy dependence term without considering the Lorentz violation term
\begin{equation}
\label{obsdelayB}
\Delta t_{\mathrm{obs}}=(1+z)\Delta t_{\mathrm{in}};
\end{equation}
\item {\bf{Model C}}: taking $\Delta t_{\rm in}$ as the general form together with the Lorentz violation term 
\begin{equation}
\label{obsdelayC}
\Delta t_{\mathrm{obs}}=\Delta t_{\mathrm{LV}}+(1+z)\Delta t_{\mathrm{in}}.
\end{equation}
Actually, Model C combines Models A and B into a unified framework~\cite{Song:2024and}.
\end{itemize}

We conduct simulations to evaluate the flexibility of Model C employed in this study. We generate three distinct datasets based on the following models:  Model A, Model B,  and Model C introduced in Ref. \cite{Song:2024and}.

We adopt the assumption that all gamma-ray bursts share similar intrinsic properties for this initial approximation. Consequently, we utilized the observed spectral parameters from GRB 221009A to generate simulated data for all 10 GRBs at varying redshifts as those of our analyzed 14+3 photons \cite{song_and_ma_ApJ}. 
The fitted spectra function at the peak interval  for GRB 221009A  can be expressed as follows~\cite{LHAASO:2023kyg}:
\begin{equation}
    \frac{dN}{dE} \propto \left(\frac{E_{\rm h,o}}{\rm TeV}\right)^{-3.006}\exp{\left(\frac{-E_{\rm h,o}}{3.14 \ {\rm TeV}}\right)}.
\end{equation}
For GRB 221009A with $z=0.151$, we can randomly draw observed high-energy photons based on the above function and then obtain the source frame energy $E_{\rm h,s} = (1+z) E_{\rm h,o}$. Assuming all GRBs have the same intrinsic spectra properties, we randomly draw 1000 high-energy photons with $E_{\rm h,s} \in [2, 20000]$ GeV for each GRB. Meanwhile, the emission time of low-energy photons are chosen as zero for all GRBs in each set. Then, the three datasets are:
\begin{itemize}
    \item \textbf{Set A}.  This dataset is generated based on Model A in Ref.~\cite{Song:2024and}  where the energy-dependent intrinsic time delay is not considered. We inject the same $a_{\rm LV} = 1/E_{\rm LV} = 2.12 \times 10^{-18} ~ {\rm GeV}^{-1}$, $\mu = -6.48$~s, and $\upsilon = 6.45$~s (we use $\upsilon$ to replace $\sigma$ in Ref.~\cite{Song:2024and}). Then, the observed time delay for a given photon with $E_{\rm h,o}$ can be obtained from Eq.~(\ref{obsdelayA}), which is equivalent to $\alpha = 0 ~ \rm{s} \cdot {\rm GeV}^{-1}$ in \textbf{Set C}. 

    \item \textbf{Set B}. This dataset is generated based on Model B in Ref.~\cite{Song:2024and} where the LV time delay is not considered. We inject the same $\alpha = 0.13 ~ \rm{s} \cdot {\rm GeV}^{-1}$, $\mu = -0.44$~s, and $\upsilon = 8.77$~s. Then, the observed time delay for a given photon with $E_{\rm h,o}$ can be obtained from Eq.~(\ref{obsdelayB}), which is equivalent to $a_{\rm LV} = 0 ~ {\rm GeV}^{-1}$ in \textbf{Set C}.

    \item \textbf{Set C}. This dataset is generated based on Model C introduced in Ref.~\cite{Song:2024and}. We inject the same $a_{\rm LV} = 3.38 \times 10^{-18} ~ {\rm GeV}^{-1}$  corresponding to $E_{\rm LV}=2.96 \times 10^{17}$ GeV, $\alpha = -0.15 ~ \rm{s} \cdot {\rm GeV}^{-1}$, $\mu = -4.98$ s, and $\upsilon = 5.67$ s. Then, the observed time delay for a given photon with $E_{\rm h,o}$ can be obtained from Eq.~(\ref{obsdelayC}).
\end{itemize}

With the exact value for photon energy and arrival time obtained through above procedure, we then consider the observational uncertainties. For the 10 GRBs in each set, we consider the previous eight GRBs observed by Fermi-LAT as discussed in Ref~\cite{Song:2024and}, and the two new GRBs, GRB 190114C and GRB 221009A observed by MAGIC and LHAASO, respectively. We consider energy resolutions to vary among different observatories, with Fermi-LAT having $\pm 10\%$ resolution \cite{Fermi-LAT:2009pgs}, MAGIC with $\pm 15\%$ 
resolution \cite{MAGIC:2019lau}, and LHAASO with a 
$\pm 20\%$ resolution \cite{LHAASO:2023lkv}.  We consider a positional uncertainty of  $\pm 5$~s for the first main significant peak of low-energy photons.

\section{Analysis methods in Bayesian framework}

In this work, we utilize the same parameter estimation methods in the Bayesian framework as in Ref.~\cite{Song:2024and}. We assume that these measurement error variables are independent and follow Gaussian distribution, and the common intrinsic time delay $\Delta t_{\mathrm{in,c}}$ follows a Gaussian distribution $\mathcal{N} \left(\mu, \upsilon^2 \right)$, \reply{where $\mu$ denotes the mean value of intrinsic time delay and $\upsilon$ quantifies the standard deviation of the emission time over the finite interval introduced in Ref.~\cite{Song:2024and} (here we substitute $\sigma$ with 
$\upsilon$ throughout our analysis)}. Meanwhile, Eq.~(\ref{obsdelay}) can be rewritten as
\begin{equation}
\frac{\Delta t_{\mathrm{obs}}}{1+z}= \Delta t_{\mathrm{in}} + \frac{\Delta t_{\rm LV}}{1+z} = \Delta t_{\mathrm{in}} + a_{\rm LV}K_1,
\end{equation}
where $a_{\rm LV} = 1/E_{\rm LV}$ and $K_1$ is,
\begin{equation}
K_1=\frac{1}{H_0}\frac{E_\mathrm{h,o}}{1+z}\int_0^z\frac{(1+z')\mathrm{d}z'}{\sqrt{\Omega_\mathrm{m}(1+z')^3+\Omega_\Lambda}}.
\end{equation}

Then, the likelihood function for Model A can be expressed as
\begin{widetext}
\begin{align*}
p \propto \exp\left[-\frac{1}{2}\sum_{j=1}^{n}\left(\frac{\left(\frac{\Delta t_{\mathrm{obs},j}}{1+z_{j}} -\mu -a_{\mathrm{LV}}K_{1,j}\right)^{2}}{\sigma^2_{y_j}+\upsilon^{2}+a_{\mathrm{LV}}^{2}\sigma_{K_{1,j}}^{2}}
+\ln\left(\sigma^2_{y_j}+\upsilon^{2}+a_{\mathrm{LV}}^{2}\sigma_{K_{1,j}}^{2}\right)\right)\right]p\left(a_{\rm LV}\right)p\left(\mu\right)p\left(\upsilon\right),
\end{align*}
\end{widetext}
where $j$ from $\{1, \dots, n\}$ denotes the $j$th observed high-energy photon and $\sigma^2_{y_j}$ represents the corresponding uncertainty for $\frac{\Delta t_{\mathrm{obs},j}}{1+z_{j}}$.
The above three parameters, $a_{\rm LV}$, $\mu$, and $\upsilon$ are assumed in flat priors
\begin{align}
    \begin{cases}
        p\left(a_{\rm LV}\right) \sim U\left[0, 30 \right] \times 10^{-18} \ \rm{GeV^{-1}}, \\
        p\left(\mu\right) \sim U\left[-30, 30 \right] \ {\rm s}  , \\
        p\left(\upsilon\right) \sim U\left[0, 30 \right] \ {\rm s}.
    \end{cases}  
\end{align}
When using Model A to analyze the mock dataset B, we mathematically expand the prior for $a_{\rm LV}$ to $U\left[-30, 30 \right] \times 10^{-18} \ \rm{GeV^{-1}}$, due to effective zero value injection.
The likelihood function for Model B is
\begin{widetext}
    \begin{equation}
    p \propto \exp\left[-\frac{1}{2}\sum_{j=1}^{n}\left(\frac{\left(\frac{\Delta t_{\mathrm{obs},j}}{1+z_{j}} -\mu - \alpha E_{(h,s),j} \right)^{2}}{\sigma^2_{y_j}+\upsilon^{2}+ \alpha^2\sigma_{E_{(h,s),j}}^2}+\ln(\sigma^2_{y_j}+\upsilon^{2} + \alpha^2\sigma_{E_{(h,s),j}}^2)\right)\right]p(\alpha)p\left(\mu\right)p\left(\upsilon\right).
    \end{equation}
\end{widetext}
The three parameters for Model B, $\alpha$, $\mu$, and $\upsilon$, are also assumed in flat priors
\begin{align}
    \begin{cases}
        p(\alpha) \sim U \left[-30, 30 \right] \ {\rm s \cdot GeV^{-1}}, \\
        p\left(\mu\right) \sim U\left[-30, 30 \right] \ {\rm s}, \\
        p\left(\upsilon\right) \sim U\left[0, 30 \right] \ {\rm s}. \\
    \end{cases}
    \label{priors}
\end{align}
The likelihood function for Model C is
\begin{widetext}
    \begin{equation}
    p \propto \exp\left[-\frac{1}{2}\sum_{j=1}^{n}\left(\frac{\left(\frac{\Delta t_{\mathrm{obs},j}}{1+z_{j}} -\mu - \alpha E_{(h,s),j} -a_{\mathrm{LV}}K_{j}\right)^{2}}{\sigma^2_{y_j}+\upsilon^{2}+ \alpha^2\sigma_{E_{(h,s),j}}^2+a_{\mathrm{LV}}^{2}\sigma_{K_{j}}^{2}}+\ln(\sigma^2_{y_j}+\upsilon^{2}+ \alpha^2\sigma_{E_{(h,s),j}}^2+a_{\mathrm{LV}}^{2}\sigma_{K_{j}}^{2})\right)\right]p\left(a_{\rm LV}\right)p(\alpha)p\left(\mu\right)p\left(\upsilon\right).
    \end{equation}
\end{widetext}
The four parameters for Model C, $a_{\rm LV}$, $\alpha$, $\mu$, and $\upsilon$ are assumed in flat priors
\begin{align}
    \begin{cases}
        p\left(a_{\rm LV}\right) \sim U\left[0, 30 \right] \times 10^{-18} \ \rm{GeV^{-1}}, \\
        p(\alpha) \sim U \left[-30, 30 \right] \ {\rm s \cdot GeV^{-1}}, \\
        p\left(\mu\right) \sim U\left[-30, 30 \right] \ {\rm s}, \\
        p\left(\upsilon\right) \sim U\left[0, 30 \right] \ {\rm s}. \\
    \end{cases}
    \label{priors}
\end{align}
When using Model C to deal with the mock dataset B, we also mathematically expand the prior for $a_{\rm LV}$ to $U\left[-30, 30 \right] \times 10^{-18} \ \rm{GeV^{-1}}$, due to the effective zero value injection.

\section{Results from mock datasets}
\begin{figure*}[h] 
	\centering 
    \begin{minipage}{0.33\textwidth}
        \centering
        \includegraphics[width=0.95\linewidth]{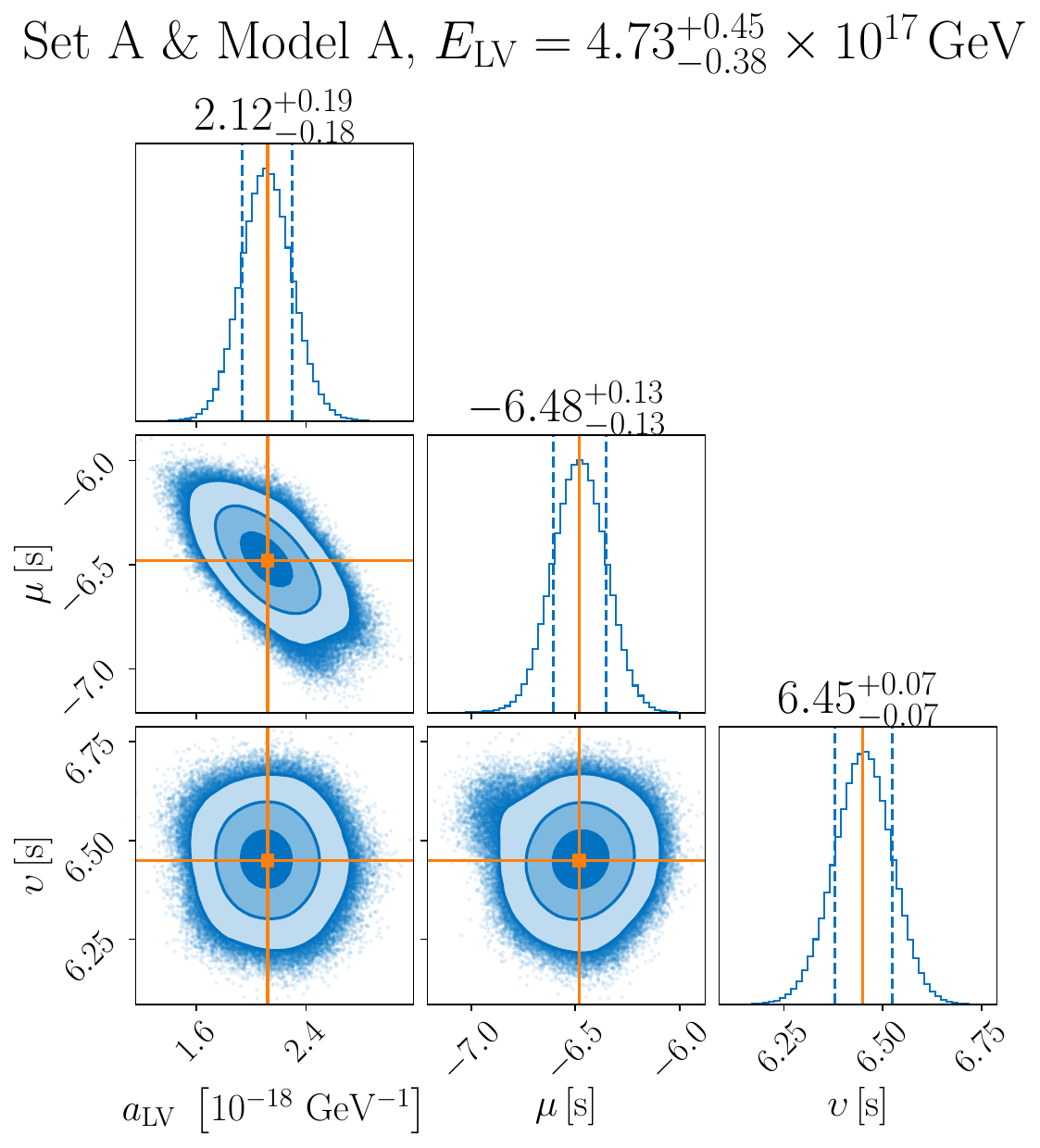}
    \end{minipage}\hfill
    \begin{minipage}{0.33\textwidth}
        \centering
        \includegraphics[width=0.95\linewidth]{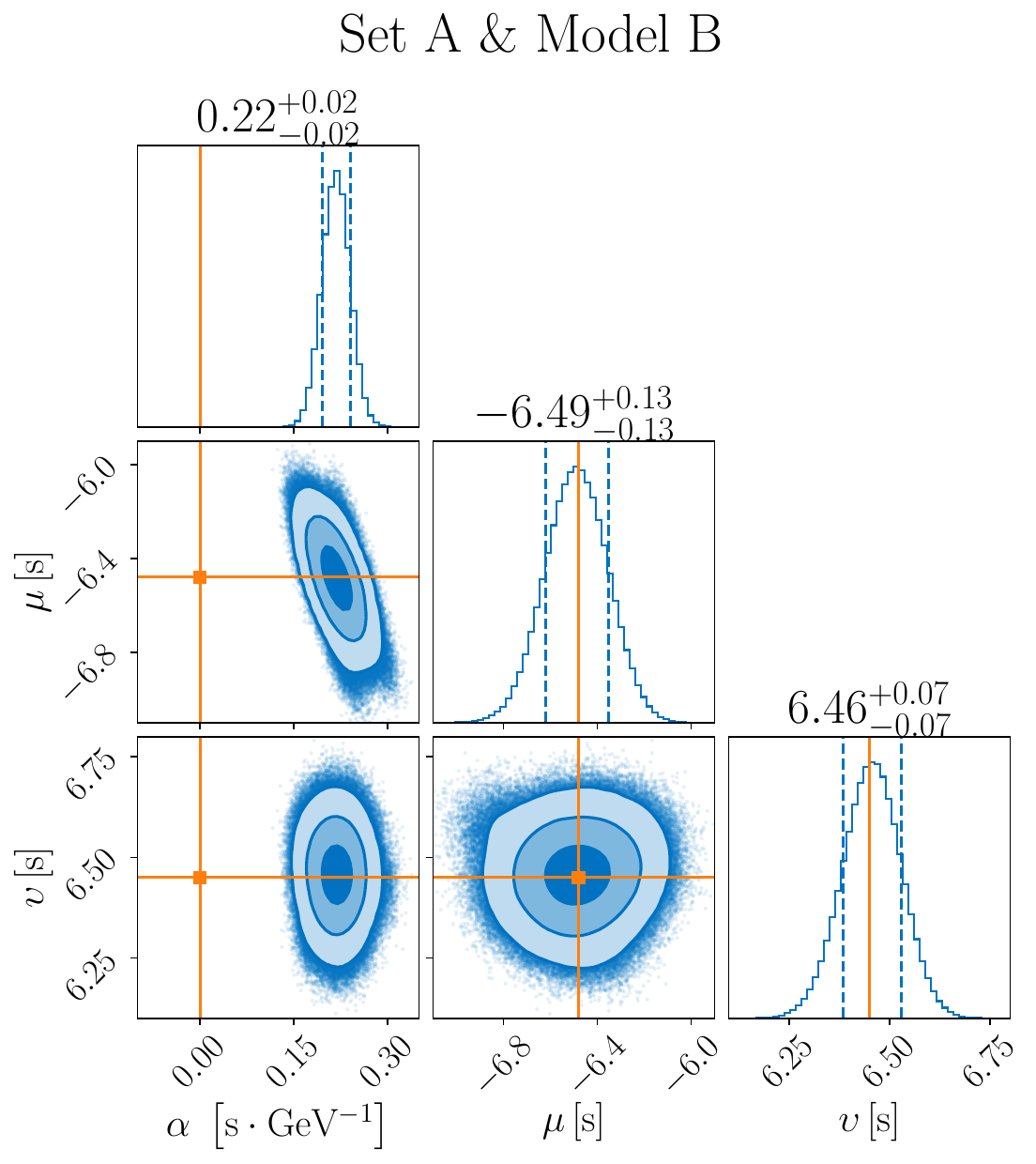} 
    \end{minipage}
    \begin{minipage}{0.33\textwidth}
        \centering
        \includegraphics[width=0.95\linewidth]{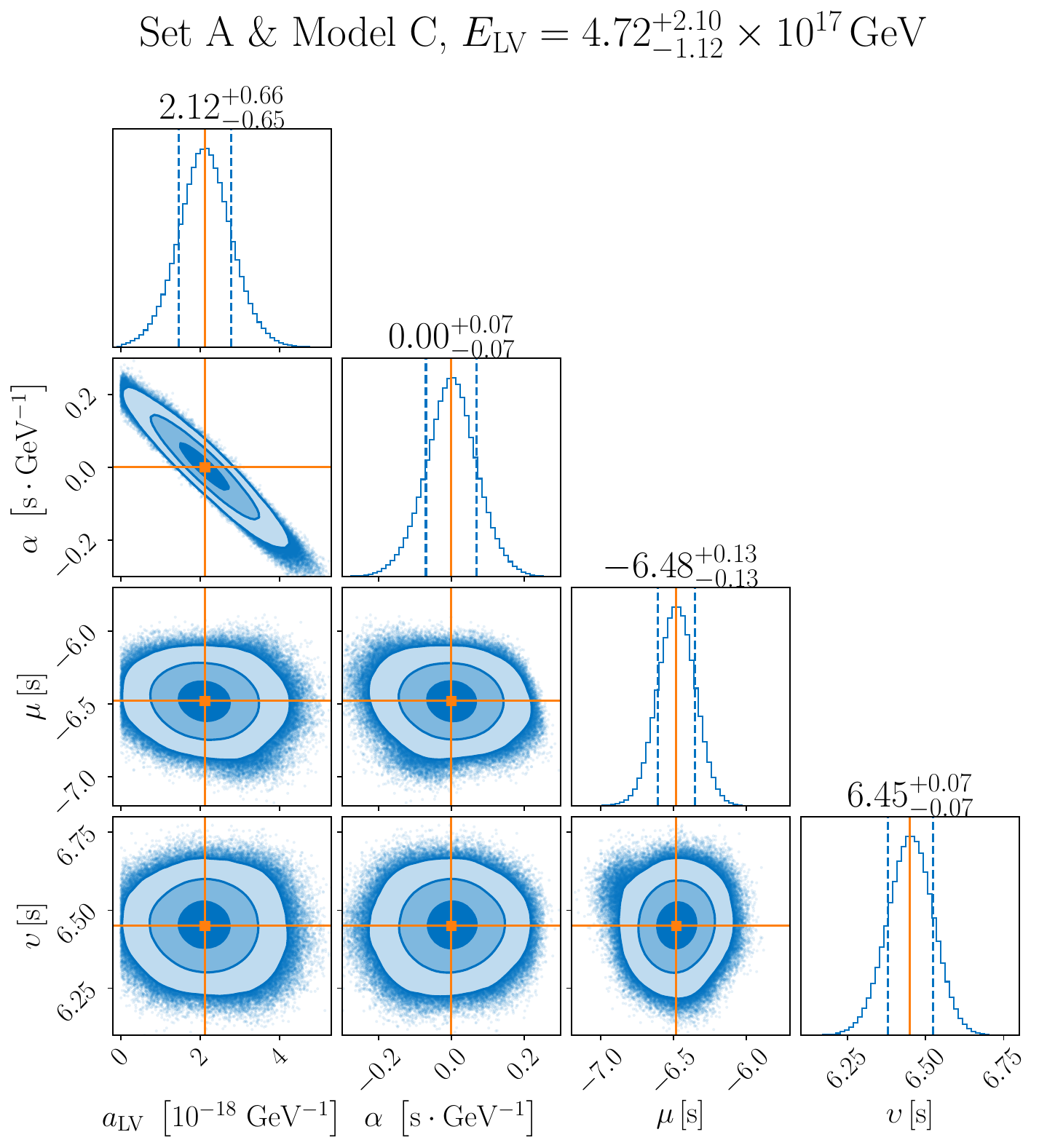} 
    \end{minipage}
    \caption{\reply{Results of examining Set A with three Models.  The 2D contours represent with different confidence levels, denoting the 1$\sigma$, 2$\sigma$, and 3$\sigma$ regions, while the dashed vertical lines indicate the  1$\sigma$ region for the 1D marginalized posterior distribution. The vertical orange lines denote the injection parameters in each mock dataset.}} 
    \label{Set_A}
\end{figure*}

\begin{figure*}[h] 
	\centering 
    \begin{minipage}{0.33\textwidth}
        \centering
        \includegraphics[width=0.95\linewidth]{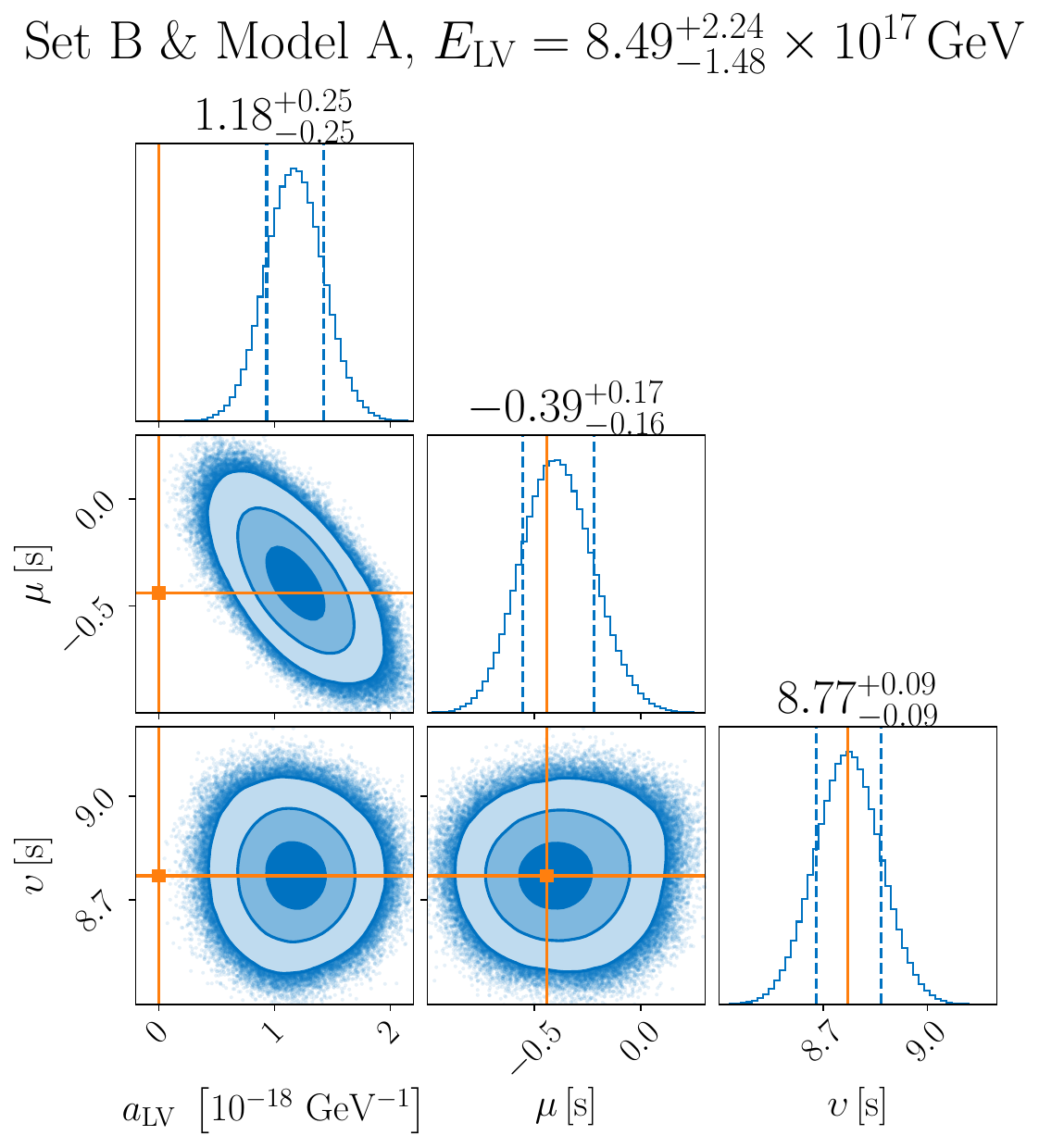}
    \end{minipage}\hfill
    \begin{minipage}{0.33\textwidth}
        \centering
        \includegraphics[width=0.95\linewidth]{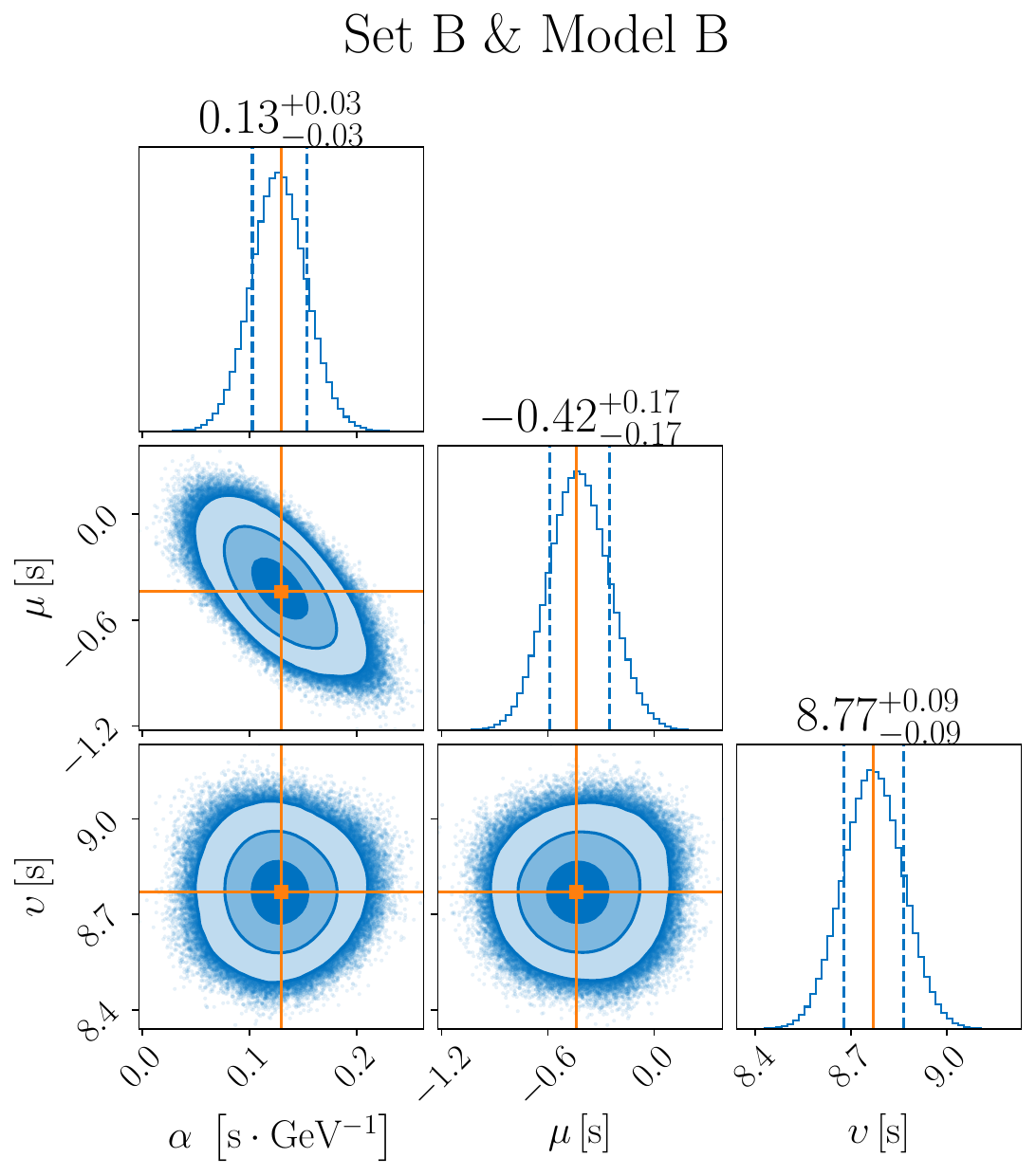} 
    \end{minipage}
    \begin{minipage}{0.33\textwidth}
        \centering
        \includegraphics[width=0.95\linewidth]{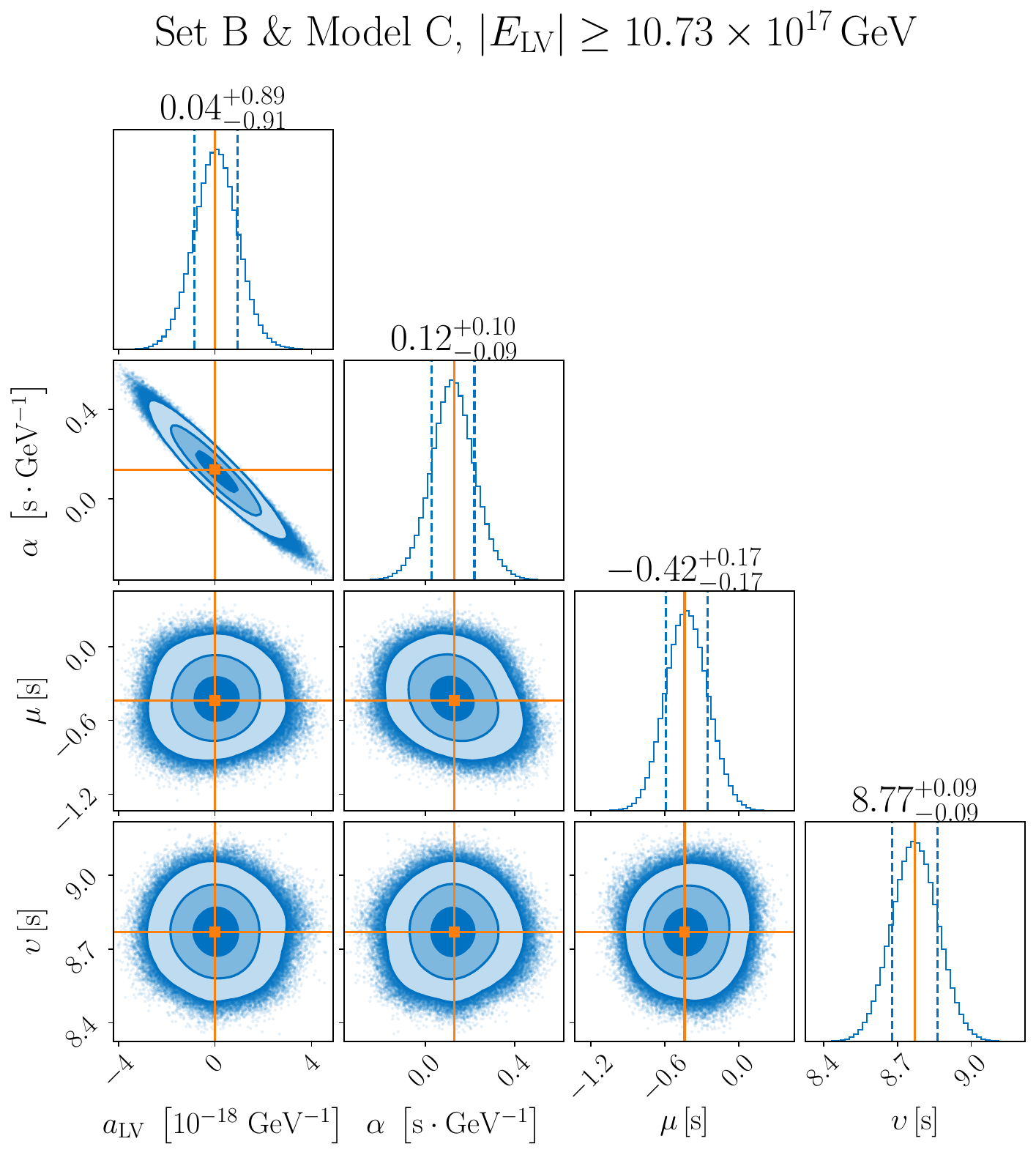} 
    \end{minipage}
    \caption{\reply{Same as Fig.~\ref{Set_A}, but for Set B.}} 
    \label{Set_B}
\end{figure*}

\begin{figure*}[h] 
	\centering 
    \begin{minipage}{0.33\textwidth}
        \centering
        \includegraphics[width=0.95\linewidth]{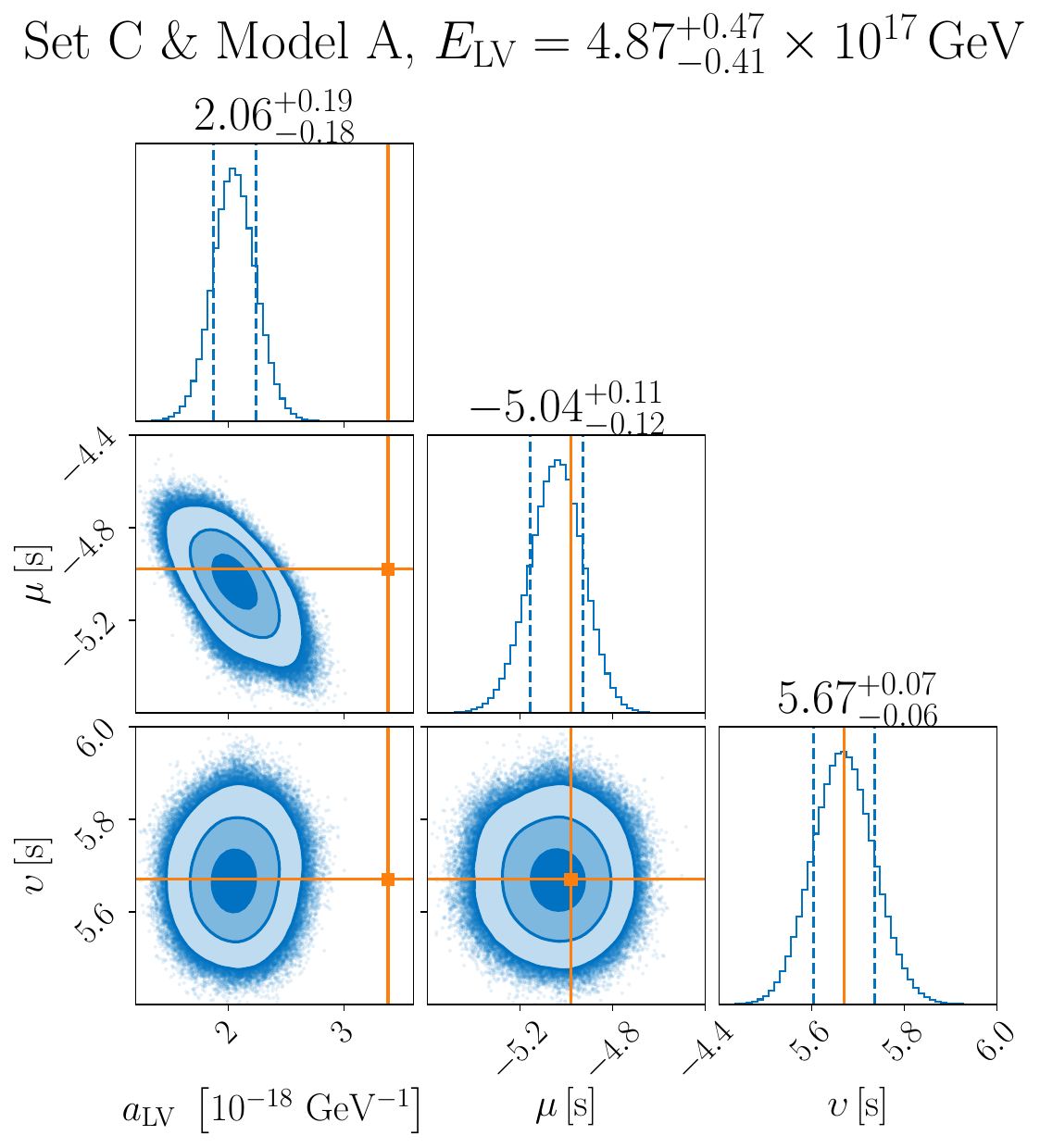}
    \end{minipage}\hfill
    \begin{minipage}{0.33\textwidth}
        \centering
        \includegraphics[width=0.95\linewidth]{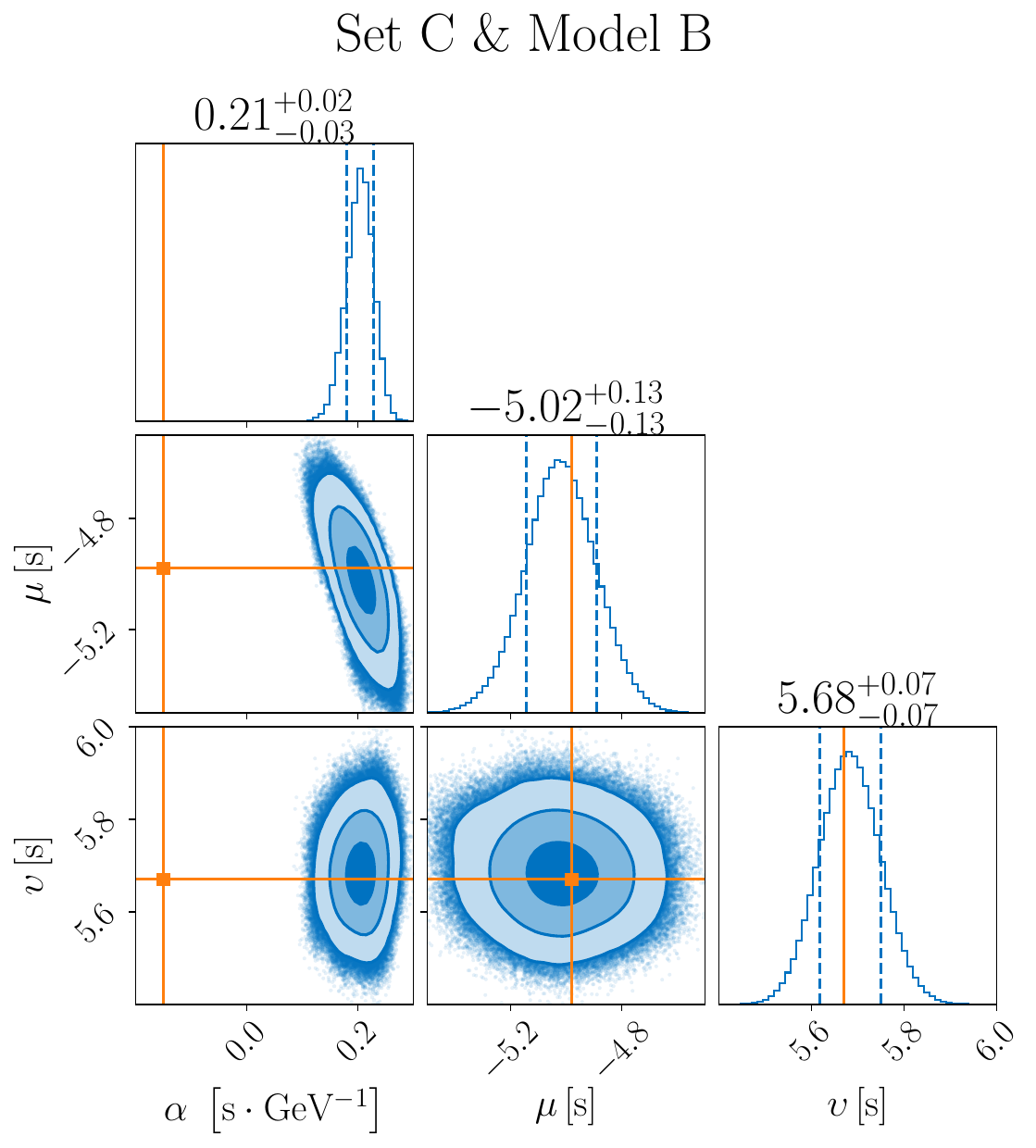} 
    \end{minipage}
    \begin{minipage}{0.33\textwidth}
        \centering
        \includegraphics[width=0.95\linewidth]{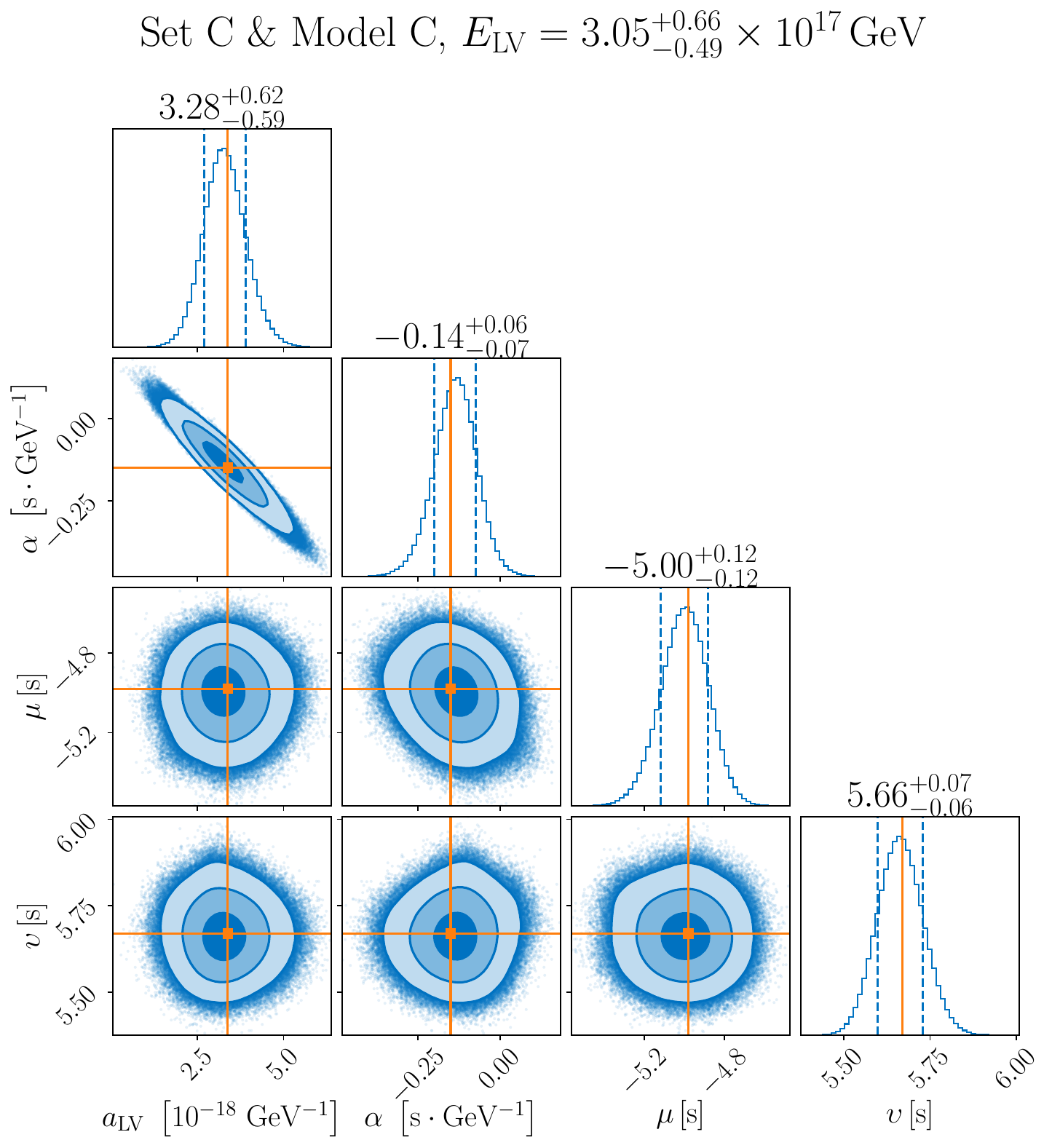} 
    \end{minipage}
    \caption{\reply{Same as Fig.~\ref{Set_A}, but for Set C.}} 
    \label{Set_C}
\end{figure*}

We systematically evaluate the performance of Models A, B, and C across three synthetic datasets (Sets A--C) using Bayesian parameter estimation. Each dataset corresponds to distinct delay-generation mechanisms: Set A (LV + constant intrinsic delays), Set B (energy-dependent intrinsic delays only), and Set C (LV + energy-dependent delays). 
It is important to note that two parts may introduce bias during the simulation: the mock data generation process and the parameter estimation process. For each dataset, we draw random values for the source-frame energy and the common intrinsic time delay for each photon. Since the random numbers generated by a computer are pseudorandom, each dataset may contain bias relative to the injected values. Additionally, performing the parameter estimation process with a limited exploration depth may also introduce bias. Thus, we repeat the mock data generation and parameter estimation processes 250 times, each time with a different noise seed. We then combine the results to mitigate the bias. Posterior distributions for each parameter are shown in Figs.~\ref{Set_A}--\ref{Set_C}, with numerical constraints summarized in Tables~\ref{param_SetA}--\ref{param_SetC}.

\begin{table*}[h]
  \centering
  \caption{\reply{Table of estimated parameters and the corresponding Lorentz violation scale $E_{\rm LV}$ with three models for Set A. The injection parameters for Set A are obtained from Ref.~\cite{Song:2024and}.}}
    \begin{tabular}{cccccc}
    \toprule
    Case & $a_{\rm LV} ~ (10^{-18} ~ {\rm GeV}^{-1})$ & $\alpha ~( \rm{s} \cdot {\rm GeV}^{-1}) $ & $\mu ~({\rm s})$ & $\upsilon ~({\rm s})$ & $E_{\rm LV} ~(10^{17}~ {\rm GeV})$ \\
    \midrule
    Injection & $2.12$ & $0$ & $-6.48$ & $6.45$ & $4.72$ \\
    Model A & $2.12^{+0.19}_{-0.18}$ & $-$  & $-6.48^{+0.13}_{-0.13}$ & $6.45^{+0.07}_{-0.07}$ & $4.73^{+0.45}_{-0.38}$\\
    Model B & $-$ & $0.22^{+0.02}_{-0.02}$ & $-6.49^{+0.13}_{-0.13}$ & $6.46^{+0.07}_{-0.07}$  & $-$\\
    Model C & $2.12^{+0.66}_{-0.66}$ & $-0.00^{+0.07}_{-0.07}$  & $-6.48^{+0.13}_{-0.13}$ &$6.45^{+0.07}_{-0.07}$ & $4.72^{+2.10}_{-1.12}$\\
    \bottomrule
    \end{tabular}
  \label{param_SetA}
\end{table*}

\begin{table*}[h]
  \centering
  \caption{\reply{Table of estimated parameters and the corresponding Lorentz violation scale $E_{\rm LV}$ with three models for Set B. The injected parameters for Set B are obtained from Ref.~\cite{Song:2024and}.} }
    \begin{tabular}{cccccc}
    \toprule
    Case & $a_{\rm LV} ~ (10^{-18} ~ {\rm GeV}^{-1})$ & $\alpha ~( \rm{s} \cdot {\rm GeV}^{-1}) $ & $\mu ~({\rm s})$ & $\upsilon ~({\rm s})$ & $E_{\rm LV} ~(10^{17}~ {\rm GeV})$ \\
    \midrule
    Injection & $0$ & $0.13$ & $-0.44$ & $8.77$ & $\infty$ \\
    Model A & $1.18^{+0.25}_{-0.25}$ & $-$  & $-0.39^{+0.17}_{-0.16}$ & $8.77^{+0.09}_{-0.09}$ & $8.49^{+2.24}_{-1.48}$\\
    Model B & $-$ & $0.13^{+0.03}_{-0.03}$ & $-0.42^{+0.17}_{-0.17}$ & $8.77^{+0.09}_{-0.09}$  & $-$\\
    Model C & $0.04^{+0.89}_{-0.91}$ & $0.12^{+0.10}_{-0.09}$  & $-0.42^{+0.17}_{-0.17}$ &$8.77^{+0.09}_{-0.09}$ & $\geq 10.73$\\
    \bottomrule
    \end{tabular}%
  \label{param_SetB}%
\end{table*}

\begin{table*}[h]
  \centering
  \caption{\reply{Table of estimated parameters and the corresponding Lorentz violation scale $E_{\rm LV}$ with three models for Set C. The injected parameters for Set C are obtained from Ref.~\cite{Song:2024and}.}}
    \begin{tabular}{cccccc}
    \toprule
    Case & $a_{\rm LV} ~ (10^{-18} ~ {\rm GeV}^{-1})$ & $\alpha ~( \rm{s} \cdot {\rm GeV}^{-1}) $ & $\mu ~({\rm s})$ & $\upsilon ~({\rm s})$ & $E_{\rm LV} ~(10^{17}~ {\rm GeV})$ \\
    \midrule
    Injection & $3.38$ & $-0.15$ & $-4.98$ & $5.67$ & $2.96$ \\
    Model A & $2.06^{+0.19}_{-0.18}$ & $-$  & $-5.04^{+0.11}_{-0.12}$ & $5.67^{+0.07}_{-0.06}$ & $4.87^{+0.47}_{-0.41}$\\
    Model B & $-$ & $0.21^{+0.02}_{-0.03}$ & $-5.02^{+0.13}_{-0.13}$ & $5.68^{+0.07}_{-0.07}$  & $-$\\
    Model C & $3.28^{+0.62}_{-0.59}$ & $-0.14^{+0.06}_{-0.07}$  & $-5.00^{+0.12}_{-0.12}$ &$5.66^{+0.07}_{-0.06}$ & $3.05^{+0.66}_{-0.49}$\\
    \bottomrule
    \end{tabular}%
  \label{param_SetC}%
\end{table*}

Fig.~\ref{Set_A} show the results for analyzing Set A with three models. Both Model A and Model C can recover the injected parameters, while Model B shows apparent bias on $\alpha$. The results for analyzing Set B with three models are shown in Fig.~\ref{Set_B}. Both Model B and Model C can recover the injected parameters, while Model A shows apparent bias on $a_{\rm LV}$. In Fig.~\ref{Set_C}, we show the results for analyzing Set C with three models. Only Model C can recover the injected parameters, while Model A shows apparent bias on $a_{\rm LV}$ and Model B shows apparent bias on $\alpha$.

More details about the results are illustrate as follows:

\paragraph*{Set A analysis (Fig.~\ref{Set_A}).}
\reply{Model A successfully recovers parameters that strictly align with the injected values for all parameters ($a_{\mathrm{LV}} = 2.12^{+0.19}_{-0.18}\times 10^{-18} \ \rm{GeV^{-1}}$, $\mu = -6.48_{-0.13}^{+0.13}~\mathrm{s}$, $\upsilon=6.45^{+0.07}_{-0.07}~\mathrm{s}$), validating its self-consistency. 
Model B, however, exhibits a \(11\sigma\) tension in \(\alpha\) (with an injected value \(\alpha = 0~\mathrm{s\cdot GeV^{-1}}\) vs.\ a recovered value \(\alpha = 0.22^{+0.02}_{-0.02}~\mathrm{s\cdot GeV^{-1}}\)), indicating a misattribution of LV-induced delays to energy-dependent astrophysical processes. Model C demonstrates an unbiased recovery across all parameters.  
The main physical parameters $a_{\mathrm{LV}}$ and $\alpha$ are strictly aligned with injected values, confirming its capacity to resolve LV-intrinsic degeneracies.}

\paragraph*{Set B analysis (Fig.~\ref{Set_B}).}
\reply{Model B yields precise constraints on energy-dependent delays ($\alpha = 0.13^{+0.03}_{-0.03}~\rm{s\cdot GeV^{-1}}$).
Model A produces a \(4.7\sigma\) overestimation of \(a_{\mathrm{LV}}\) (with an injected value $a_{\mathrm{LV}} = 0\ \rm{GeV^{-1}}$ vs. a recovered value $a_{\mathrm{LV}} = 1.18^{+0.25}_{-0.25}\times 10^{-18}\ \rm{GeV^{-1}}$), demonstrating catastrophic failure when neglecting energy-dependent astrophysical terms. Model C once again achieves an accurate  recovery of the injected parameters, resolving all terms simultaneously.}

\paragraph*{Set C analysis (Fig.~\ref{Set_C}).}
\reply{Only Model C recovers the joint LV+astrophysical parameters within statistical uncertainties (\(a_{\mathrm{LV}} = 3.28^{+0.62}_{-0.59}\times 10^{-18}\ \rm{GeV^{-1}}\), \(\alpha = -0.14^{+0.06}_{-0.07}~\mathrm{s\cdot GeV^{-1}}\)). Model A catastrophically misestimates \(a_{\mathrm{LV}}\) with \(6.9\sigma\) tension (with an injected value \(a_{\mathrm{LV}}\) = \(3.38\times 10^{-18}\) vs. a recovered value  \(a_{\mathrm{LV}}=2.06^{+0.19}_{-0.18}\times 10^{-18}\ \rm{GeV^{-1}}\)) by absorbing energy-dependent delays into LV terms, 
while Model B fails to capture energy-dependent intrinsic emission feature (with the injected value $\alpha =-0.15~\mathrm{s\cdot GeV^{-1}}$, \(12\sigma\) tension from the recovered value \(\alpha = 0.21^{+0.02}_{-0.03}~\mathrm{s\cdot GeV^{-1}}\)).}\\

These results establish three critical findings:
\begin{itemize}
    \item \textbf{Model C universality}: Recovers parameters with sub-\(\sigma\) accuracy across all datasets (right plots of Figs.~\ref{Set_A}, \ref{Set_B}, \ref{Set_C}). 
    The main physical parameters $a_{\rm LV}$ and $\alpha$ can be recovered within 1$\sigma$ for all three datasets. 
    
    \item \textbf{Model A/B limitations}: Exhibit \(>4.7\sigma\) biases when applied to datasets generated under alternative mechanisms for key parameters (Tables~\ref{param_SetA} and \ref{param_SetB}), proving inadequate for multimechanism analyses.
    \item \textbf{Energy dependence necessity}: The \(\alpha\) parameter's \(>6.9\sigma\) tension in fitting Set C with Model A conclusively demonstrates that intrinsic delays cannot be treated as energy-independent constants.
\end{itemize}

The collective evidence positions Model C as the sole framework capable of disentangling LV signatures from astrophysical systematics in current and next-generation GRB observations.

\section{Results from Realistic Datasets: Multi-GeV to TeV Constraints with Fermi-LAT, MAGIC, and LHAASO Observations}
\label{sec:realdata}

\subsection{Parameter degeneracy in single-burst analyses}

In this section, we address a critical aspect of Model C when only photon event data from a single gamma-ray burst (GRB) is available. The relationship expressed in Eq.~(\ref{obsdelayC}) can be reformulated as
\[
\Delta t_{\rm obs}/(1+z) = E_{\rm h,s}[a_{\rm LV} f(z) + \alpha] + \Delta t_{\rm in,c},
\]
where \( f(z) \) is a function that depends solely on the redshift \( z \) of the source and the assumed cosmological model. For each source with a fixed \( z \), \( f(z) \) can be accurately determined. This implies that the parameters \( a_{\rm LV} \) and \( \alpha \) can be effectively replaced by a single parameter \( C = a_{\rm LV} f(z) + \alpha \) for photons originating from a single GRB. Consequently, \( a_{\rm LV} \) and \( \alpha \) exhibit complete degeneracy.

The key point lies in the parameter \( C = a_{\rm LV} f(z) + \alpha \). As discussed in Ref.~\cite{Song:2024and}, for any individual GRB event, the redshift \( z \) is known and remains constant for photons emitted from that GRB, allowing \( f(z) \) to be treated as a constant. As a result, the parameters \( a_{\rm LV} \) and \( \alpha \) become indistinguishable. However, when multiple GRB events are analyzed, this degeneracy can be broken, as the relationship between \( a_{\rm LV} f(z) \) and \( \alpha \) will vary across events from different GRBs. Therefore, \( a_{\rm LV} \) and \( \alpha \) cannot be consolidated into a single parameter \( C = a_{\rm LV} f(z) + \alpha \).

The degeneracy between Lorentz violation time delays and intrinsic time delays when analyzing a single GRB event (i.e., photons from the same GRB) can be resolved in two ways: by analyzing multiple GRB events collectively (i.e., considering photons from different GRBs) or by using one of the parameters \( a_{\rm LV} \) or \( \alpha \) obtained from a multi-GRB analysis as an input when examining photon events from a single GRB.

\subsection{Multi-GRB analysis with Fermi-LAT data}

Ref.~\cite{Song:2024and} conducted analyses using three distinct models (Models A, B, and C) to evaluate the Fermi-LAT dataset comprising 14 events with intrinsic energies exceeding 30 GeV from eight GRBs. The findings revealed different scenarios regarding the intrinsic emission of high-energy photons and the scales of Lorentz violation. Model A aligns with prior research \cite{Xu:2016zxi, Xu:2016zsa, Zhu:2021pml}, yielding consistent results with a Lorentz invariance violation energy scale of \( E_{\rm LV} \simeq 4.71 \times 10^{17} \) GeV. In contrast, Model B suggests that a positive source energy-correlated intrinsic time delay term may account for the observed time delay phenomenon in GRBs. Conversely, Model C indicates the presence of a negative source energy-correlated intrinsic time delay term alongside a reduced Lorentz violation scale of \( E_{\rm LV} \simeq 2.96 \times 10^{17} \) GeV. These significant distinctions highlight the necessity for additional GRB data to effectively disentangle and assess the contributions from LV-induced time delays and intrinsic time delays.

\subsection{Incorporating TeV photons: MAGIC and LHAASO constraints}

\begin{figure*}[t] 
	\centering 
    \begin{minipage}{0.33\textwidth}
        \centering
        \includegraphics[width=0.95\linewidth]{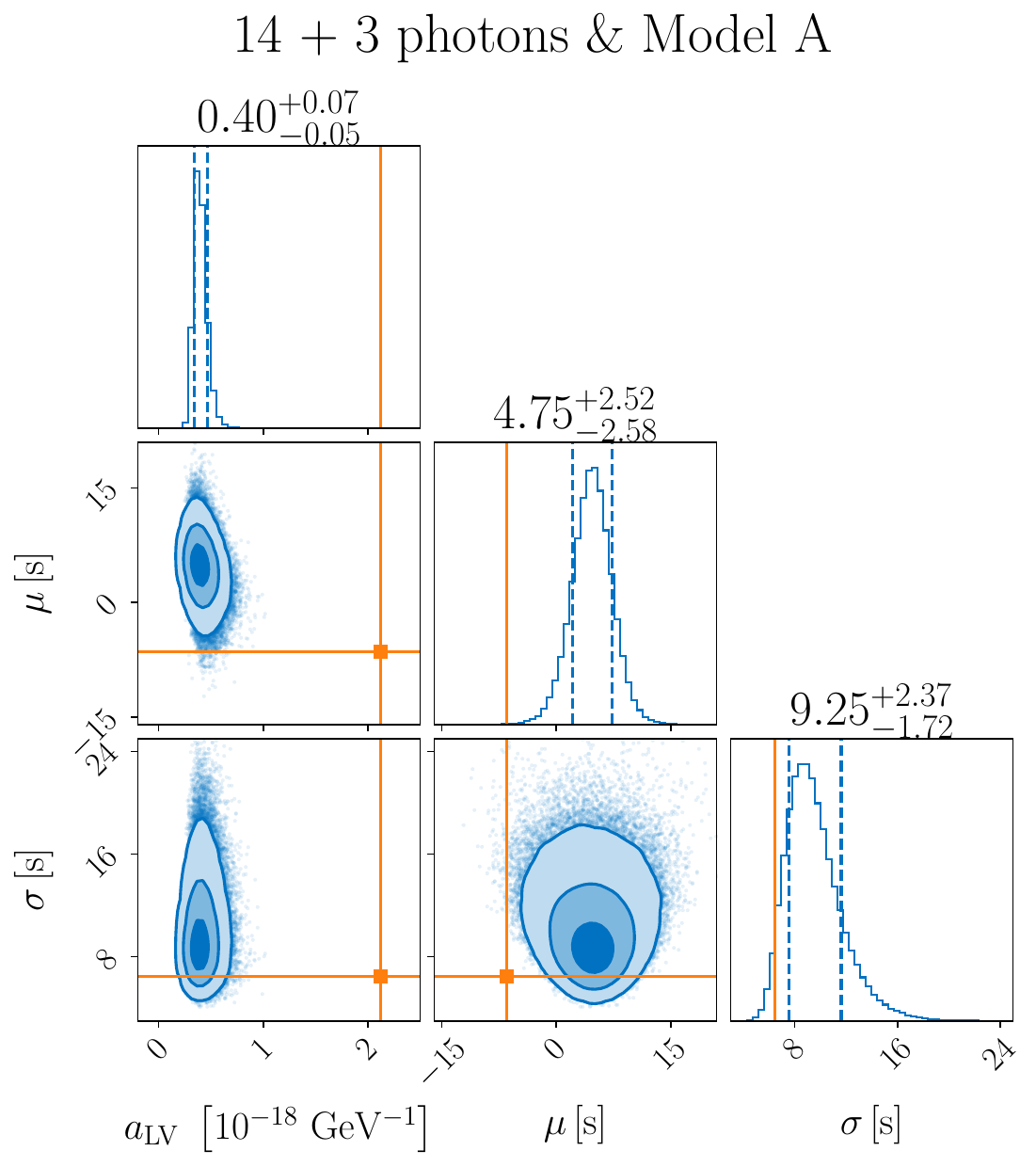}
    \end{minipage}\hfill
    \begin{minipage}{0.33\textwidth}
        \centering
        \includegraphics[width=0.95\linewidth]{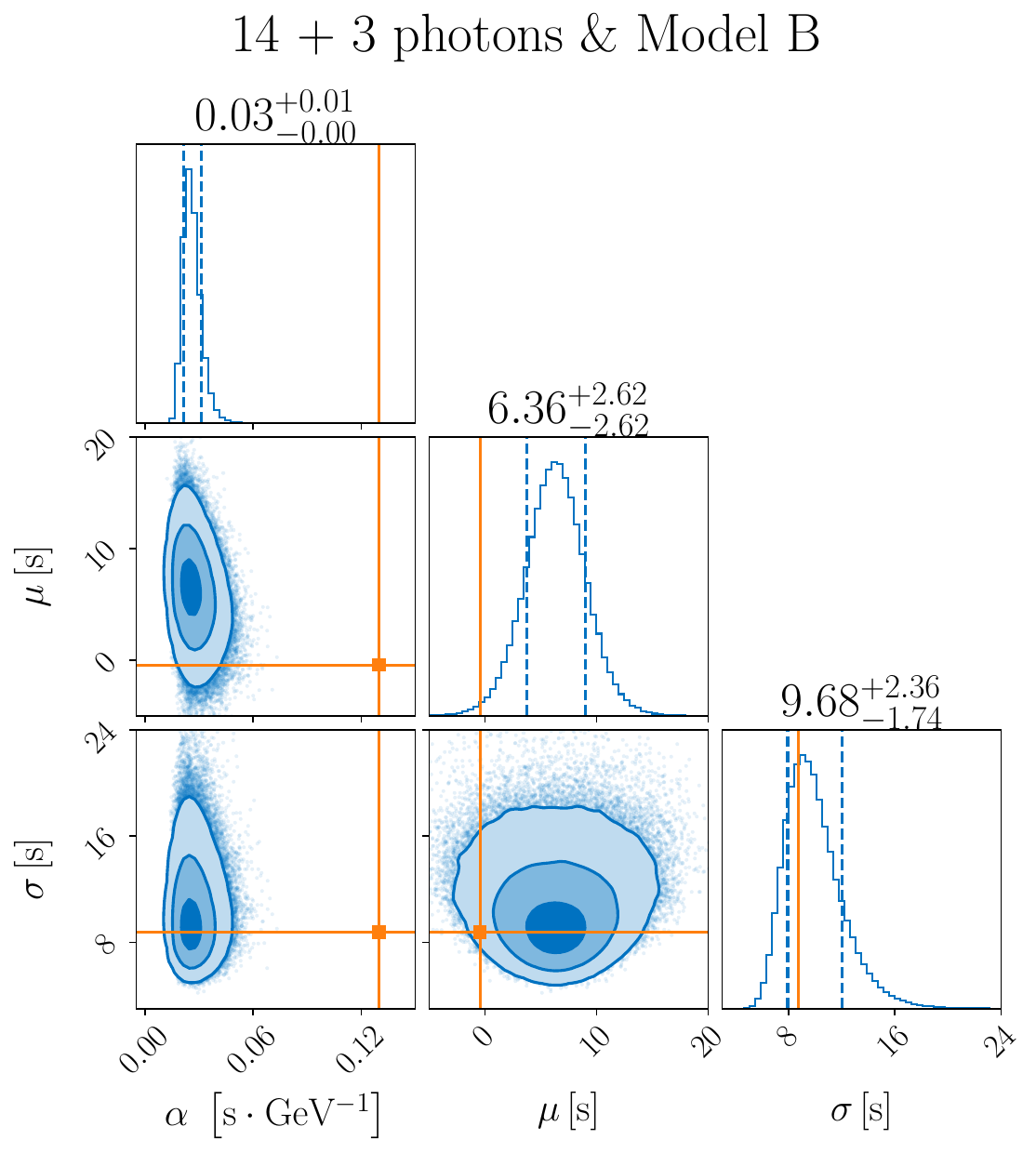} 
    \end{minipage}
    \begin{minipage}{0.33\textwidth}
        \centering
        \includegraphics[width=0.95\linewidth]{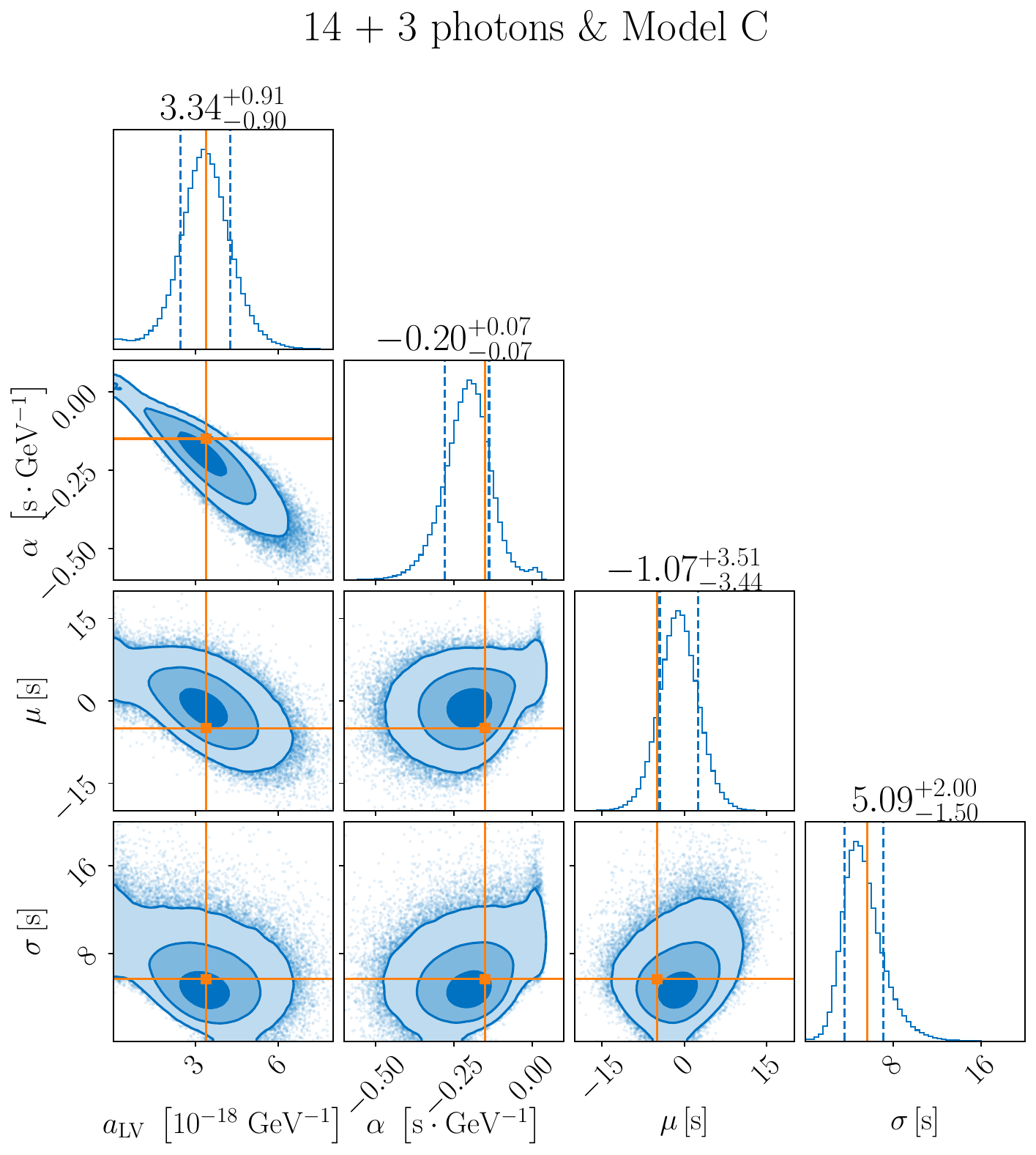} 
    \end{minipage}
    \caption{Examining three models with 14 + 3 photons from Ref.~\cite{Song:2025qej}. The orange lines in each subfigure denotes the inference results with 14 multi-GeV photons from Ref.~\cite{Song:2024and}, which are also the injected parameters for Set A, Set B, and Set C, respectively.} 
    \label{13+3_photons}
\end{figure*}

In Ref.~\cite{Song:2025qej}, a further analysis was conducted with the inclusion of three notable GRB photons recorded by different observatories: the 99.3 GeV photon from GRB 221009A observed by Fermi-LAT, the 1.07 TeV photon from GRB 190114C detected by the Major Atmospheric Gamma Imaging Cherenkov (MAGIC) telescope, and the 12.2 TeV photon from GRB 221009A observed by the Large High Altitude Air Shower Observatory (LHAASO), in addition to the Fermi-LAT dataset of 14 events from eight GRBs. The results presented in Ref.~\cite{Song:2025qej} (as shown in the left plot of Fig.~\ref{13+3_photons}) indicate that Model A fails to provide a consistent framework to reconcile the remarkable MAGIC and LHAASO TeV photons with the Fermi-LAT multi-GeV photons under the same Lorentz violation scale \( E_{\rm LV} \).

For the results associated with Model B in Ref.~\cite{Song:2025qej}, one can conclude that the observed time delays of high-energy photons are primarily attributed to intrinsic emission time delays occurring at the GRB source frames, permitting a larger Lorentz violation scale \( E_{\rm LV} \) that is comparable to or even exceeds the Planck scale \( E_{\rm p} \simeq 1.22 \times 10^{19} \) GeV. However, the parameter
$\alpha$ differs significantly with those from the Fermi-LAT dataset
by including the MAGIC TeV photon and LHAASO multi-TeV photon, showing the inconsistency between the two datasets with Model B, as shown in the middle plot of Fig.~\ref{13+3_photons}. 

The results obtained from Model C are particularly intriguing, as they reveal a remarkable consistency in the parameters across various combinations of datasets, as illustrated in the right plot of Fig.~\ref{13+3_photons}.
More detailed illustrations and in-depth analyses on the results from Model~C can be found in Ref.~\cite{song_and_ma_ApJ}.
Notably, our findings align with those of previous studies \cite{Xu:2016zsa,Xu:2016zxi,Song:2024and}, reinforcing the hypothesis of Lorentz violation characterized by a violation scale of \( E_{\rm LV} \sim 3 \times 10^{17} \) GeV. This suggests that high-energy photons are emitted prior to low-energy photons within the frames of the GRB sources. Such consistency across different datasets not only strengthens the case for Lorentz violation but also emphasizes the importance of considering the emission sequence of photons in understanding the underlying physics of gamma-ray bursts.

\subsection{Implications for LV detection frameworks}

The above results suggest that Model A and Model B fail to consistently explain the Fermi-LAT dataset in conjunction with the MAGIC TeV photon and the LHAASO multi-TeV photon. However, Model C proves to be a viable candidate for a consistent descriptions of all mentioned datasets, of course further investigation is still required to determine its ability to explain additional data in the future. This will facilitate the formulation of reliable conclusions regarding the intrinsic emission mechanisms of GRBs and Lorentz violation scales inferred from cosmic GRB photons.

Model C has one more free parameter (i.e., $a_{\rm{LV}}$) in comparison
with Model B. 
From a technical perspective, increasing the number of free parameters does not always enhance the performance of a physics model. Fortunately, one can utilize the Akaike information criterion for model selection to determine the model that best fits the data \cite{akaike1981likelihood, Biesiada:2009zz}.  An analysis~\cite{Song:2025qej} along this line indicates that Model C exhibits the best performance on analyzing the combined Fermi-LAT+MAGIC+LHAASO dataset, whereas Model B demonstrates the poorest performance.

The persistent $\alpha < 0$ preference in Model C suggests fundamental emission mechanism differences—high-energy photons originate from distinct jet regions or earlier acceleration phases. A recent analysis~\cite{Liu:2024qbt} of LHAASO observations \cite{LHAASO:2023lkv,LHAASO:2023kyg} tentatively supports this idea through precursor signals.

Future facilities like Cherenkov Telescope Array (CTA) \cite{CTAConsortium:2010umy} will revolutionize LV studies through:
\begin{itemize}
    \item $\sim$1000 GRB events with $z > 2$ probes;
    \item millisecond timing at 0.1-100~TeV energies;
    \item 3D or multi-D parameter reconstruction ($E_{\mathrm{LV}}, \alpha, z$).
\end{itemize}

This establishes Model C as the current preferred framework while highlighting TeV photons as crucial probes of both quantum gravity and compact object astrophysics.

\section{Conclusions}

We perform Monte Carlo simulations using synthetic datasets comprising 10 GRBs matched in redshift to observed counterparts. Adopting the intrinsic spectral properties of GRB 221009A measured by LHAASO-WCDA \cite{LHAASO:2023kyg} as a universal template, we generate 1000 high-energy photons per GRB in the source frame, spanning energies $E_{\rm high,s} \in [2, 20000]$~GeV. Three distinct mock datasets based on Models A, B and C are constructed by embedding time delays governed by
\begin{equation}
\label{model}
\frac{\Delta t_{\mathrm{obs}}}{1+z} = a_{\mathrm{LV}}K_1 + \alpha E_{\mathrm{h,s}} + \Delta t_{\mathrm{in,c}}, \quad \Delta t_{\mathrm{in,c}} \sim \mathcal{N}(\mu, \upsilon^2),
\end{equation}
where parameters ($a_{\mathrm{LV}},\alpha, \mu, \upsilon$) are injected from real-data posterior distributions.

Bayesian analysis reveals that Model C—jointly incorporating Lorentz violation (LV) delays and energy-dependent intrinsic emission terms—successfully recovers injected parameters across all synthetic datasets, achieving subpercent biases in $E_{\rm LV}$ and $\alpha$ estimates. Conversely, Models A (LV + constant delay) and B (energy-dependent delay only) exhibit catastrophic failures when applied to datasets generated under opposing mechanisms, with parameter mismatches exceeding $\sim5\sigma$ significance in TeV-energy regimes. This dichotomy highlights the unique capacity of Model C to disentangle LV signatures from astrophysical systematics.

To validate our framework, we analyze recent findings from studies~\cite{Song:2024and,song_and_ma_ApJ, Song:2025qej} that focus on realistic datasets comprising 14 Fermi-LAT multi-GeV photons from eight gamma-ray bursts. Additionally, we incorporate a 99.3 GeV photon from GRB 221009A observed by Fermi-LAT, a TeV photon from GRB 190114C detected by the MAGIC telescope, and a multi-TeV photon from GRB 221009A recorded by LHAASO. While Models A and B produce unphysical constraints on Lorentz violation or result in null detections, Model C robustly identifies a subluminal Lorentz violation characterized by an energy scale of $E_{\rm LV} \simeq 3 \times 10^{17}$ GeV, which is consistent with earlier constraints. Our results clearly demonstrate that energy-dependent intrinsic delays—previously overlooked in prior studies—are essential for reconciling LV signals across different energy bands. This highlights the importance of incorporating energy-dependent effects in future analyses to enhance the understanding of Lorentz violation in high-energy astrophysics.

This systematic validation establishes Model C as the definitive framework for future LV searches with high-energy observatories (e.g., LHAASO and CTA), ensuring reliable discrimination between quantum-gravity-induced dispersion and source astrophysics. \\

\emph{\textbf{Acknowledgments}.---}
This work is supported by National Natural Science Foundation of China under Grants No.~12335006 and No.~12075003.  This work is also supported by High-performance Computing Platform of Peking University.

\bibliography{scibib}

\begin{thebibliography}{10}
\expandafter\ifx\csname url\endcsname\relax
  \def\url#1{\texttt{#1}}\fi
\expandafter\ifx\csname urlprefix\endcsname\relax\def\urlprefix{URL }\fi
\expandafter\ifx\csname href\endcsname\relax
  \def\href#1#2{#2} \def\path#1{#1}\fi

\bibitem{Amelino-Camelia:1996bln}
G.~Amelino-Camelia, J.~R. Ellis, N.~E. Mavromatos, D.~V. Nanopoulos,
  \href{https://doi.org/10.1142/S0217751X97000566}{Distance measurement and
  wave dispersion in a Liouville string approach to quantum gravity}, Int. J.
  Mod. Phys. A 12 (1997) 607--624.
\newblock \href {http://arxiv.org/abs/hep-th/9605211}
  {\path{arXiv:hep-th/9605211}}, \href
  {https://doi.org/10.1142/S0217751X97000566}
  {\path{doi:10.1142/S0217751X97000566}}.

\bibitem{Amelino-Camelia:1997ieq}
G.~Amelino-Camelia, J.~R. Ellis, N.~E. Mavromatos, D.~V. Nanopoulos, S.~Sarkar,
  \href{https://doi.org/10.1038/31647}{Tests of quantum gravity from
  observations of gamma-ray bursts}, Nature 393 (1998) 763--765.
\newblock \href {http://arxiv.org/abs/astro-ph/9712103}
  {\path{arXiv:astro-ph/9712103}}, \href {https://doi.org/10.1038/31647}
  {\path{doi:10.1038/31647}}.

\bibitem{Ellis:1999rz}
J.~R. Ellis, N.~E. Mavromatos, D.~V. Nanopoulos,
  \href{https://doi.org/10.1023/A:1026720723556}{Search for quantum gravity},
  Gen. Rel. Grav. 31 (1999) 1257--1262.
\newblock \href {http://arxiv.org/abs/gr-qc/9905048}
  {\path{arXiv:gr-qc/9905048}}, \href {https://doi.org/10.1023/A:1026720723556}
  {\path{doi:10.1023/A:1026720723556}}.

\bibitem{Ellis:1999uh}
J.~R. Ellis, N.~E. Mavromatos, D.~V. Nanopoulos,
  \href{https://doi.org/10.1023/A:1001852601248}{Quantum gravitational
  diffusion and stochastic fluctuations in the velocity of light}, Gen. Rel.
  Grav. 32 (2000) 127--144.
\newblock \href {http://arxiv.org/abs/gr-qc/9904068}
  {\path{arXiv:gr-qc/9904068}}, \href {https://doi.org/10.1023/A:1001852601248}
  {\path{doi:10.1023/A:1001852601248}}.

\bibitem{Ellis:2008gg}
J.~R. Ellis, N.~E. Mavromatos, D.~V. Nanopoulos,
  \href{https://doi.org/10.1016/j.physletb.2008.06.029}{Derivation of a Vacuum
  Refractive Index in a Stringy Space-Time Foam Model}, Phys. Lett. B 665
  (2008) 412--417.
\newblock \href {http://arxiv.org/abs/0804.3566} {\path{arXiv:0804.3566}},
  \href {https://doi.org/10.1016/j.physletb.2008.06.029}
  {\path{doi:10.1016/j.physletb.2008.06.029}}.

\bibitem{Li:2009tt}
T.~Li, N.~E. Mavromatos, D.~V. Nanopoulos, D.~Xie,
  \href{https://doi.org/10.1016/j.physletb.2009.07.062}{Time Delays of Strings
  in D-particle Backgrounds and Vacuum Refractive Indices}, Phys. Lett. B 679
  (2009) 407--413.
\newblock \href {http://arxiv.org/abs/0903.1303} {\path{arXiv:0903.1303}},
  \href {https://doi.org/10.1016/j.physletb.2009.07.062}
  {\path{doi:10.1016/j.physletb.2009.07.062}}.

\bibitem{Li:2021gah}
C.~Li, B.-Q. Ma, \href{https://doi.org/10.1016/j.physletb.2021.136443}{Light
  speed variation in a string theory model for space-time foam}, Phys. Lett. B
  819 (2021) 136443.
\newblock \href {http://arxiv.org/abs/2105.06151} {\path{arXiv:2105.06151}},
  \href {https://doi.org/10.1016/j.physletb.2021.136443}
  {\path{doi:10.1016/j.physletb.2021.136443}}.

\bibitem{Li:2021eza}
C.~Li, B.-Q. Ma, \href{https://doi.org/10.1016/j.rinp.2021.104380}{Light speed
  variation with brane/string-inspired space-time foam}, Results Phys. 26
  (2021) 104380.
\newblock \href {http://arxiv.org/abs/2109.07096} {\path{arXiv:2109.07096}},
  \href {https://doi.org/10.1016/j.rinp.2021.104380}
  {\path{doi:10.1016/j.rinp.2021.104380}}.

\bibitem{Gambini:1998it}
R.~Gambini, J.~Pullin,
  \href{https://doi.org/10.1103/PhysRevD.59.124021}{Nonstandard optics from
  quantum space-time}, Phys. Rev. D 59 (1999) 124021.
\newblock \href {http://arxiv.org/abs/gr-qc/9809038}
  {\path{arXiv:gr-qc/9809038}}, \href
  {https://doi.org/10.1103/PhysRevD.59.124021}
  {\path{doi:10.1103/PhysRevD.59.124021}}.

\bibitem{Alfaro:1999wd}
J.~Alfaro, H.~A. Morales-Tecotl, L.~F. Urrutia,
  \href{https://doi.org/10.1103/PhysRevLett.84.2318}{Quantum gravity
  corrections to neutrino propagation}, Phys. Rev. Lett. 84 (2000) 2318--2321.
\newblock \href {http://arxiv.org/abs/gr-qc/9909079}
  {\path{arXiv:gr-qc/9909079}}, \href
  {https://doi.org/10.1103/PhysRevLett.84.2318}
  {\path{doi:10.1103/PhysRevLett.84.2318}}.

\bibitem{Li:2022szn}
H.~Li, B.-Q. Ma, \href{https://doi.org/10.1016/j.physletb.2022.137613}{Speed
  variations of cosmic photons and neutrinos from loop quantum gravity}, Phys.
  Lett. B 836 (2023) 137613.
\newblock \href {http://arxiv.org/abs/2212.04220} {\path{arXiv:2212.04220}},
  \href {https://doi.org/10.1016/j.physletb.2022.137613}
  {\path{doi:10.1016/j.physletb.2022.137613}}.

\bibitem{Amelino-Camelia:2002cqb}
G.~Amelino-Camelia, \href{https://doi.org/10.1038/418034a}{Doubly special
  relativity}, Nature 418 (2002) 34--35.
\newblock \href {http://arxiv.org/abs/gr-qc/0207049}
  {\path{arXiv:gr-qc/0207049}}, \href {https://doi.org/10.1038/418034a}
  {\path{doi:10.1038/418034a}}.

\bibitem{Amelino-Camelia:2002uql}
G.~Amelino-Camelia, \href{https://doi.org/10.1142/S021827180200302X}{Doubly
  special relativity: First results and key open problems}, Int. J. Mod. Phys.
  D 11 (2002) 1643.
\newblock \href {http://arxiv.org/abs/gr-qc/0210063}
  {\path{arXiv:gr-qc/0210063}}, \href
  {https://doi.org/10.1142/S021827180200302X}
  {\path{doi:10.1142/S021827180200302X}}.

\bibitem{Amelino-Camelia:2000stu}
G.~Amelino-Camelia, \href{https://doi.org/10.1142/S0218271802001330}{Relativity
  in space-times with short distance structure governed by an observer
  independent (Planckian) length scale}, Int. J. Mod. Phys. D 11 (2002) 35--60.
\newblock \href {http://arxiv.org/abs/gr-qc/0012051}
  {\path{arXiv:gr-qc/0012051}}, \href
  {https://doi.org/10.1142/S0218271802001330}
  {\path{doi:10.1142/S0218271802001330}}.

\bibitem{He:2022gyk}
P.~He, B.-Q. Ma, \href{https://doi.org/10.3390/universe8060323}{Lorentz
  Symmetry Violation of Cosmic Photons}, Universe 8~(6) (2022) 323.
\newblock \href {http://arxiv.org/abs/2206.08180} {\path{arXiv:2206.08180}},
  \href {https://doi.org/10.3390/universe8060323}
  {\path{doi:10.3390/universe8060323}}.

\bibitem{AlvesBatista:2023wqm}
R.~Alves~Batista, et~al., {White paper and roadmap for quantum gravity
  phenomenology in the multi-messenger era}, Class. Quant. Grav. 42~(3) (2025)
  032001.
\newblock \href {http://arxiv.org/abs/2312.00409} {\path{arXiv:2312.00409}},
  \href {https://doi.org/10.1088/1361-6382/ad605a}
  {\path{doi:10.1088/1361-6382/ad605a}}.

\bibitem{Xiao:2009xe}
Z.~Xiao, B.-Q. Ma,
  \href{https://doi.org/10.1103/PhysRevD.80.116005}{Constraints on Lorentz
  invariance violation from gamma-ray burst GRB090510}, Phys. Rev. D 80 (2009)
  116005.
\newblock \href {http://arxiv.org/abs/0909.4927} {\path{arXiv:0909.4927}},
  \href {https://doi.org/10.1103/PhysRevD.80.116005}
  {\path{doi:10.1103/PhysRevD.80.116005}}.

\bibitem{Jacob:2008bw}
U.~Jacob, T.~Piran,
  \href{https://doi.org/10.1088/1475-7516/2008/01/031}{Lorentz-violation-induced
  arrival delays of cosmological particles}, JCAP 01 (2008) 031.
\newblock \href {http://arxiv.org/abs/0712.2170} {\path{arXiv:0712.2170}},
  \href {https://doi.org/10.1088/1475-7516/2008/01/031}
  {\path{doi:10.1088/1475-7516/2008/01/031}}.

\bibitem{Zhu:2022blp}
J.~Zhu, B.-Q. Ma,
  \href{https://doi.org/10.1103/PhysRevD.105.124069}{Lorentz-violation-induced
  arrival time delay of astroparticles in Finsler spacetime}, Phys. Rev. D
  105~(12) (2022) 124069.
\newblock \href {http://arxiv.org/abs/2206.07616} {\path{arXiv:2206.07616}},
  \href {https://doi.org/10.1103/PhysRevD.105.124069}
  {\path{doi:10.1103/PhysRevD.105.124069}}.

\bibitem{Shao:2009bv}
L.~Shao, Z.~Xiao, B.-Q. Ma,
  \href{https://doi.org/10.1016/j.astropartphys.2010.03.003}{Lorentz violation
  from cosmological objects with very high energy photon emissions}, Astropart.
  Phys. 33 (2010) 312--315.
\newblock \href {http://arxiv.org/abs/0911.2276} {\path{arXiv:0911.2276}},
  \href {https://doi.org/10.1016/j.astropartphys.2010.03.003}
  {\path{doi:10.1016/j.astropartphys.2010.03.003}}.

\bibitem{Zhang:2014wpb}
S.~Zhang, B.-Q. Ma,
  \href{https://doi.org/10.1016/j.astropartphys.2014.04.008}{Lorentz violation
  from gamma-ray bursts}, Astropart. Phys. 61 (2015) 108--112.
\newblock \href {http://arxiv.org/abs/1406.4568} {\path{arXiv:1406.4568}},
  \href {https://doi.org/10.1016/j.astropartphys.2014.04.008}
  {\path{doi:10.1016/j.astropartphys.2014.04.008}}.

\bibitem{Xu:2016zxi}
H.~Xu, B.-Q. Ma,
  \href{https://doi.org/10.1016/j.astropartphys.2016.05.008}{Light speed
  variation from gamma-ray bursts}, Astropart. Phys. 82 (2016) 72--76.
\newblock \href {http://arxiv.org/abs/1607.03203} {\path{arXiv:1607.03203}},
  \href {https://doi.org/10.1016/j.astropartphys.2016.05.008}
  {\path{doi:10.1016/j.astropartphys.2016.05.008}}.

\bibitem{Xu:2016zsa}
H.~Xu, B.-Q. Ma, \href{https://doi.org/10.1016/j.physletb.2016.07.044}{Light
  speed variation from gamma ray burst GRB 160509A}, Phys. Lett. B 760 (2016)
  602--604.
\newblock \href {http://arxiv.org/abs/1607.08043} {\path{arXiv:1607.08043}},
  \href {https://doi.org/10.1016/j.physletb.2016.07.044}
  {\path{doi:10.1016/j.physletb.2016.07.044}}.

\bibitem{Liu:2018qrg}
Y.~Liu, B.-Q. Ma, \href{https://doi.org/10.1140/epjc/s10052-018-6294-y}{Light
  speed variation from gamma ray bursts: criteria for low energy photons}, Eur.
  Phys. J. C 78~(10) (2018) 825.
\newblock \href {http://arxiv.org/abs/1810.00636} {\path{arXiv:1810.00636}},
  \href {https://doi.org/10.1140/epjc/s10052-018-6294-y}
  {\path{doi:10.1140/epjc/s10052-018-6294-y}}.

\bibitem{Zhu:2021pml}
J.~Zhu, B.-Q. Ma,
  \href{https://doi.org/10.1016/j.physletb.2021.136518}{Pre-burst events of
  gamma-ray bursts with light speed variation}, Phys. Lett. B 820 (2021)
  136518.
\newblock \href {http://arxiv.org/abs/2108.05804} {\path{arXiv:2108.05804}},
  \href {https://doi.org/10.1016/j.physletb.2021.136518}
  {\path{doi:10.1016/j.physletb.2021.136518}}.

\bibitem{Zhu:2021wtw}
J.~Zhu, B.-Q. Ma,
  \href{https://doi.org/10.1016/j.physletb.2021.136546}{Pre-burst neutrinos of
  gamma-ray bursters accompanied by high-energy photons}, Phys. Lett. B 820
  (2021) 136546.
\newblock \href {http://arxiv.org/abs/2108.08425} {\path{arXiv:2108.08425}},
  \href {https://doi.org/10.1016/j.physletb.2021.136546}
  {\path{doi:10.1016/j.physletb.2021.136546}}.

\bibitem{Zhu:2022usw}
J.~Zhu, B.-Q. Ma, \href{https://doi.org/10.1088/1361-6471/accebb}{Light speed
  variation from GRB 221009A}, J. Phys. G 50~(6) (2023) 06LT01.
\newblock \href {http://arxiv.org/abs/2210.11376} {\path{arXiv:2210.11376}},
  \href {https://doi.org/10.1088/1361-6471/accebb}
  {\path{doi:10.1088/1361-6471/accebb}}.

\bibitem{Song:2024and}
H.~Song, B.-Q. Ma, {Energy-dependent intrinsic time delay of gamma-ray bursts
  on testing Lorentz invariance violation}, Phys. Lett. B 856 (2024) 138951.
\newblock \href {http://arxiv.org/abs/2408.14719} {\path{arXiv:2408.14719}},
  \href {https://doi.org/10.1016/j.physletb.2024.138951}
  {\path{doi:10.1016/j.physletb.2024.138951}}.

\bibitem{Lesage:2023vvj}
S.~Lesage, et~al., \href{https://doi.org/10.3847/2041-8213/ace5b4}{Fermi-GBM
  Discovery of GRB 221009A: An Extraordinarily Bright GRB from Onset to
  Afterglow}, Astrophys. J. Lett. 952~(2) (2023) L42.
\newblock \href {http://arxiv.org/abs/2303.14172} {\path{arXiv:2303.14172}},
  \href {https://doi.org/10.3847/2041-8213/ace5b4}
  {\path{doi:10.3847/2041-8213/ace5b4}}.

\bibitem{LHAASO:2023lkv}
Z.~Cao, et~al., \href{https://doi.org/10.1126/sciadv.adj2778}{Very high energy
  gamma-ray emission beyond 10 TeV from GRB 221009A}, Sci. Adv. 9~(46) (2023)
  adj2778.
\newblock \href {http://arxiv.org/abs/2310.08845} {\path{arXiv:2310.08845}},
  \href {https://doi.org/10.1126/sciadv.adj2778}
  {\path{doi:10.1126/sciadv.adj2778}}.

\bibitem{MAGIC:2019lau}
V.~A. Acciari, et~al., {Teraelectronvolt emission from the $\gamma$-ray burst
  GRB 190114C}, Nature 575~(7783) (2019) 455--458.
\newblock \href {http://arxiv.org/abs/2006.07249} {\path{arXiv:2006.07249}},
  \href {https://doi.org/10.1038/s41586-019-1750-x}
  {\path{doi:10.1038/s41586-019-1750-x}}.

\bibitem{song_and_ma_ApJ}
H.~Song, B.-Q. Ma, {Lorentz invariance violation from gamma-ray bursts},
  Astrophys. J. 983~(1) (2025) 9.
\newblock \href {http://arxiv.org/abs/2504.00918} {\path{arXiv:2504.00918}},
  \href {https://doi.org/10.3847/1538-4357/adb8d4}
  {\path{doi:10.3847/1538-4357/adb8d4}}.

\bibitem{Song:2025qej}
H.~Song, B.-Q. Ma, {Examining Lorentz invariance violation with three
  remarkable GRB photons}, Phys. Dark Univ. 47 (2025) 101808.
\newblock \href {http://arxiv.org/abs/2504.14295} {\path{arXiv:2504.14295}},
  \href {https://doi.org/10.1016/j.dark.2025.101808}
  {\path{doi:10.1016/j.dark.2025.101808}}.

\bibitem{LHAASO:2023kyg}
Z.~Cao, et~al., {A tera\textendash{}electron volt afterglow from a narrow jet
  in an extremely bright gamma-ray burst}, Science 380~(6652) (2023) adg9328.
\newblock \href {http://arxiv.org/abs/2306.06372} {\path{arXiv:2306.06372}},
  \href {https://doi.org/10.1126/science.adg9328}
  {\path{doi:10.1126/science.adg9328}}.

\bibitem{Fermi-LAT:2009pgs}
A.~A. Abdo, et~al., \href{https://doi.org/10.1103/PhysRevLett.103.251101}{Fermi
  Large Area Telescope Measurements of the Diffuse Gamma-Ray Emission at
  Intermediate Galactic Latitudes}, Phys. Rev. Lett. 103 (2009) 251101.
\newblock \href {http://arxiv.org/abs/0912.0973} {\path{arXiv:0912.0973}},
  \href {https://doi.org/10.1103/PhysRevLett.103.251101}
  {\path{doi:10.1103/PhysRevLett.103.251101}}.

\bibitem{akaike1981likelihood}
H.~Akaike, Likelihood of a model and information criteria, Journal of
  Econometrics 16~(1) (1981) 3--14.

\bibitem{Biesiada:2009zz}
M.~Biesiada, A.~Piorkowska, {Lorentz invariance violation-induced time delays
  in GRBs in different cosmological models}, Class. Quant. Grav. 26 (2009)
  125007.
\newblock \href {http://arxiv.org/abs/1008.2615} {\path{arXiv:1008.2615}},
  \href {https://doi.org/10.1088/0264-9381/26/12/125007}
  {\path{doi:10.1088/0264-9381/26/12/125007}}.

\bibitem{Liu:2024qbt}
Q.~Liu, H.~Song, B.-Q. Ma, {Direct Evidence for Preburst Stage of Gamma-Ray
  Burst from GRB\,221009A Data}, Res. Notes AAS 8~(10) (2024) 263.
\newblock \href {http://arxiv.org/abs/2410.12243} {\path{arXiv:2410.12243}},
  \href {https://doi.org/10.3847/2515-5172/ad85e7}
  {\path{doi:10.3847/2515-5172/ad85e7}}.

\bibitem{CTAConsortium:2010umy}
M.~Actis, et~al., {Design concepts for the Cherenkov Telescope Array CTA: An
  advanced facility for ground-based high-energy gamma-ray astronomy}, Exper.
  Astron. 32 (2011) 193--316.
\newblock \href {http://arxiv.org/abs/1008.3703} {\path{arXiv:1008.3703}},
  \href {https://doi.org/10.1007/s10686-011-9247-0}
  {\path{doi:10.1007/s10686-011-9247-0}}.

\end{thebibliography}

\end{document}